\documentclass[traditabstract,a4paper]{aa} 

\usepackage{txfonts}
\usepackage{graphicx}
\usepackage{natbib}
\usepackage[dvips]{epsfig}
\usepackage{float}
\usepackage{subfigure}

\begin{document}
   \title{Numerical simulations of stellar jets and comparison between 
          synthetic and observed maps: clues to the launch mechanism}

   \author{
          F. Rubini \inst{1}
          \and
          L. Maurri\inst{1}
          \and
          G. Inghirami\inst{1,2}
          \and
          F. Bacciotti\inst{3}
          \and
          L. Del Zanna\inst{1,2,3}
          }

   \institute{
             Dipartimento di Fisica e Astronomia, Universit\`a di Firenze,
             Via G. Sansone 1, I-50019, Sesto F.~no (Firenze), Italy
             \and INFN - Sezione di Firenze, 
             Via G. Sansone 1, I-50019, Sesto F.~no (Firenze), Italy
             \and INAF - Osservatorio Astrofisico di Arcetri, 
              Largo E. Fermi 5, I-50125, Firenze, Italy
                }

   \date{Received / Accepted}

\abstract
{
High angular resolution spectra obtained with the Hubble Space Telescope 
Imaging Spectrograph (HST/STIS) provide rich morphological and 
kinematical information about the stellar jet phenomenon, which allows 
us to test theoretical models efficiently.
In this work, numerical simulations of stellar jets in the propagation
region are executed with the PLUTO code, by adopting inflow conditions
that arise from former numerical simulations of magnetized
outflows, accelerated by the disk-wind mechanism in the launching region.
By matching the two regions, information about the magneto-centrifugal 
accelerating mechanism underlying a given astrophysical object can be 
extrapolated by comparing synthetic and observed 
position-velocity diagrams (PVDs).
We show that quite different jets, like those from the young 
T Tauri stars DG-Tau and RW-Aur, 
may originate from the same disk-wind model for 
different configurations of the magnetic field at the disk surface.
This result supports the idea that all the observed jets may 
be generated by the same mechanism.
}


\titlerunning{Numerical simulations of stellar jets} 

\authorrunning{F. Rubini et al.}

   \maketitle
%

\section{Introduction}
\label{introduction}

%
The jet phenomenon appears to be very robust and ubiquitous in nature, 
as collimated outflows are 
seen on a large variety of spatial scales and masses of the central source, 
from active galactic nuclei to compact objects, and to forming stars.
Since the jet properties scale with the depth of the gravitational 
potential well of the central object, it is widely viewed that
jet formation may rely on an universal mechanism.
Observations can only provide the elements necessary to test the validity 
of the proposed theories for stellar jets, however, 
because of the proximity of star formation regions and the abundance 
of emitted spectral lines. 

Stellar jets are found in association with accretion disks, 
and are observed at all stages of the  formation process,
from the protostar phase to the dissipation of the disk itself.
Jets are associated with all kinds of stellar masses and are
believed to play a fundamental role in the formation process, 
as they may regulate the 
extraction of the excess angular momentum from the star-disk system,  
allowing the accretion of matter onto the central source.
For this reason, in recent years, jets from young stellar objects 
have been the target  of many observational campaigns at different 
wavelengths \citep{Bally07}. In particular, high angular resolution images and spectra taken 
with the Hubble Space Telescope (HST) at optical wavelengths have provided 
unprecedented information on the morphology and kinematics of these systems,
with data resolved spectrally and spatially both along and across 
the flow \citep[see, e.g.,][]{Bacciotti00, Bacciotti02a, Hartigan07, Coffey07}.
One of the most interesting results of these observations is the radial
velocity shift between two sides of a jet with respect to the axis
\citep{Bacciotti02b, Woitas05, Coffey08}. If interpreted as
jet rotation, this result would confirm the idea that jets carry away 
angular momentum from the  accretion disk in line with the magneto-hydrodynamic (MHD) 
models, and the jet generation process would be due to
the combination of magnetic and centrifugal forces
\citep{Blandford-Payne82,Pudritz92,Ferreira97}.

According to the magneto-centrifugal theory,  
the wind material is launched 
along the magnetic surfaces attached to the rotating  
star and disk.  The winds are then collimated magnetically 
just a few AUs above the star/disk by the action of a self-generated strong 
toroidal magnetic field. In the process, the excess angular momentum is extracted 
from the system and carried away with the jet, and the matter is allowed to accrete 
onto the star, whose rotation is also slowed down. 
The theory is elegant and very general, 
and within its framework  various models have been proposed that differ for the 
region from which the disk is accelerated: stellar winds from the star surface
\citep{Sauty99}, the X-wind, originating from the inner 
edge of the accretion disk \citep{Shu00}, or the disk-wind, 
where particles are launched from an extended region of the magnetized, 
Keplerian disk \citep{Pudritz07}.

Unfortunately, observations cannot directly test the magneto-centrifugal
launch mechanism yet because of the small scale involved (a few AUs at most). 
In fact, neither the optics of Hubble Space Telescope (HST) nor the ground telescopes 
equipped with adaptive optics can reach an angular resolution higher than $0.1''$,
equivalent to $\sim 12-14$~AUs for the nearest star formation regions.
In addition, the dusty formation cloud can hide the launching region in younger systems, 
which favors the observations of jets from more evolved, cleared  T-Tauri stars.
Finally, in both cases the interpretation of the jet spectra is 
made difficult by projection effects along the line of sight.
Nevertheless, valuable observational studies at high resolution and 
in different wavelengths have successfully investigated
the properties of the jets in the acceleration region immediately downstream from
the collimation zone, between $\sim 10$ and 100 AU from the star, and have provided 
important constraints to the launch process \citep[e.g.,] []{Bacciotti00,Woitas02,
Melnikov09,Coffey08,Takami04,Pyo03,Pyo06,Hartigan07,Agra-Amboage11}.
The only way to connect the observed region to the launch zone, however, 
is to create a logical link via numerical studies.

Numerical simulations are a very powerful tool to go beyond 
the limits encountered by observations, and to probe
theoretical models or interpret observational data.
Nevertheless, running simulations that include both
the launching region, a few AU wide, and
the propagation region, extending up to hundreds or thousands AU
from the star-disk system, is still a goal beyond the present possibilities. 
In fact, the values of the physical quantities differ by  many orders of magnitude 
in the two zones, and this introduces severe  numerical difficulties. 
For this reason the two regions are, usually, investigated separately.
So, the jet launching is investigated with numerical simulations 
based on magneto-centrifugal models 
\citep[e.g.,][]{Pudritz06,Zanni07,Meliani06,Romanova09,Tzeferacos13},
while the jet propagation is studied by running simulations of   
a supersonic collimated jet flowing beyond an ideal nozzle placed at some distance from the disk 
\citep[e.g.,][]{Rubini07,Bonito10}.

In this paper, the problem has been overcome with a procedure that operates a 
matching of the values of the quantities derived from the simulations in the  two zones.
This is done by imposing that the supersonic jet evolves in the propagation region 
from a formerly accelerated outflow arising from the launching and collimation region.
The general acceleration-propagation matching procedure developed by the authors, 
(hereafter APM procedure, see Sect.~\ref{setup}) goes through the following three steps:

\begin{enumerate}

\item The determination of the quantities in the launching-collimation region 
bounded by the rotating disk
has been addressed by referring to numerical  works already in the literature, and in particular 
the  study of \citet{Pudritz06} (hereafter P06). 
This paper provides axisymmetric numerical solutions 
of outflows accelerated magneto-centrifugally by an underlying accretion disk.
The nondimensional solutions 
described there have been specialized to the cases of interest to us by
introducing proper scale factors (stellar mass, width of the launching region, gas parameters,
etc\dots)  derived from observational studies of the real cases considered. 

\item We use the solutions derived in this way  to provide the 
inflow conditions for numerical simulations of propagating jets 
flowing from a nozzle, ideally placed at the 
border of the accelerating region. The propagation is then followed by 
axisymmetric simulations in ($r, z$) cylindrical coordinates, 
which evolve the outflow up to distances of the 
order of hundreds of AUs, i.e., the regions observed at high angular resolution.
In this model, the point ($z=0, r=0$) 
represents the star position, and the nozzle is placed
at the distance $z_\mathrm{nozzle}$ from the disk,  
representing  the border of the launching region simulated in P06.
The final result of this step is a set of maps of the physical quantities  
in the jet  meridional plane produced at the end of the run.

\item The gas parameters calculated in the propagation region are entered 
as input values in a code developed by us (Optical Telescope Simulator, OTS)
to generate images and synthetic position-velocity diagrams of the line 
emission as they would 
be produced by a given instrument working at optical wavelengths. 
The synthetic maps are produced by calculating the 
emissivity in each cell on the basis of the physical quantities 
obtained in step 2, and then integrating through the body of the jet 
and along the line of sight. 
The aim of this step is to compare the emission produced by the numerical 
jet to the emission observed in reality to get feedback 
on the validity of the assumptions made for the 
launch mechanism in the first step. 

\end{enumerate}

In this paper, we focus  on the initial section of jets from cleared T-Tauri stars, 
and in particular on the bright 5 arcseconds  of the outflows from  
DG-Tau \citep{Bacciotti00, Bacciotti02a, Bacciotti02b,Coffey08},
and from RW-Aur \citep{Woitas02,Woitas05, Melnikov09}, 
for which a big wealth of data obtained at $0.1''$ with HST is available.
In the third step, therefore, we mimic the properties of the 
Space Telescope Imaging Spectrograph (STIS) on-board HST,
using in the OTS code the same spatial 
and spectral resolution and slit setting of the real  instrument.
As we show in the following, this procedure demonstrates to be a powerful tool  
to set useful constraints on the launch process, 
and to proceed toward a complete understanding of 
the role of collimated outflows in the formation of a stellar system.    

The paper is organized as follows. The indications on kinematic and 
morphological 
properties of DG-Tau and RW-Aur jets  provided by observations are 
summarized in Sect.~\ref{observations}. Sect.~\ref{setup} describes  
the general setup procedure, including the choice of the physical 
parameters (subsection ~\ref{inflow}) and details about the 
profiles of the quantities at the inlet (subsection ~\ref{profiles}). 
In particular, Sect.~\ref{ppcode} contains a short description of the 
post-processing OTS code designed to return synthetic PVDs. 
Our results are described in Sect.~\ref{results}, and a discussion is offered 
in Sect.~\ref{conclusions}, together with our main conclusions.
In the Appendix, we discuss the validity of our results with
an additional test case simulation.

\section{Observations of the DG-Tau and RW-Aur jets} 
\label{observations}

The jet associated with DG-Tau, named HH~158,  was one of the 
first Harbig-Haro objects discovered \citep{Mundt83, Solf93}, 
and the jet-like nature was definitely confirmed by \cite{Lavalley97}. 
Its inclination with respect to the 
line of sight is about 38$^{\circ}$ \citep{Eisloffel98}, and in its 
first arcseconds the jet appears as a 
series of luminous opening bubbles, as suggested first in the images 
formed in bright optical forbidden
lines in \citet{Lavalley97,Lavalley-Fouquet00, Dougados02}. 
Further downstream, the object HH-702 
at $\sim 11''$ away from the source may be part 
of the same outflow \citep{McGroarty07}, indicating that the bright microjet 
close to the star flows in the wake of a previous emission episode.
The observations in the [\ion{Fe}{ii}]~$\lambda$1.644~$\mu$m line 
discussed in \citet{Pyo03} reveal two distinct radial velocity components
in the blue-shifted lobe, and detect the redshifted counterflow whose emission
features suggest the presence of an optically thick circumstellar disk of
$\sim$ 140 AU in radius.
A  warm molecular wind component thermalized at 2000~K
has also been detected \citep{Takami04}. 

The jet has been widely studied at high resolution 
with HST/STIS, as described in  \citet{Bacciotti00,Bacciotti02a,Coffey07,Coffey08}.
Recently,  \cite{Maurri14} (hereafter MA14), 
provided a complete set of position-velocity diagrams of the 
surface brightness in this jet, and maps of physical quantities derived 
with spectral diagnostic techniques.
Finally, multiepoch observations in the X-ray domain 
\citep{Guedel08} show a rich phenomenology. In particular,
luminous moving outer knots appear to fade compatibly with cooling models,
while closer to the star there is a component showing a steady source of  
soft X-Rays.
These observations provide a huge amount of information that must
be taken into account in the set-up of the numerics.
In particular, the PVD maps  in MA14 constitute the testbed 
of our simulations for this jet. 

\begin{figure*}[!htb]
 \resizebox{\textwidth}{!}{\includegraphics[angle=180]
{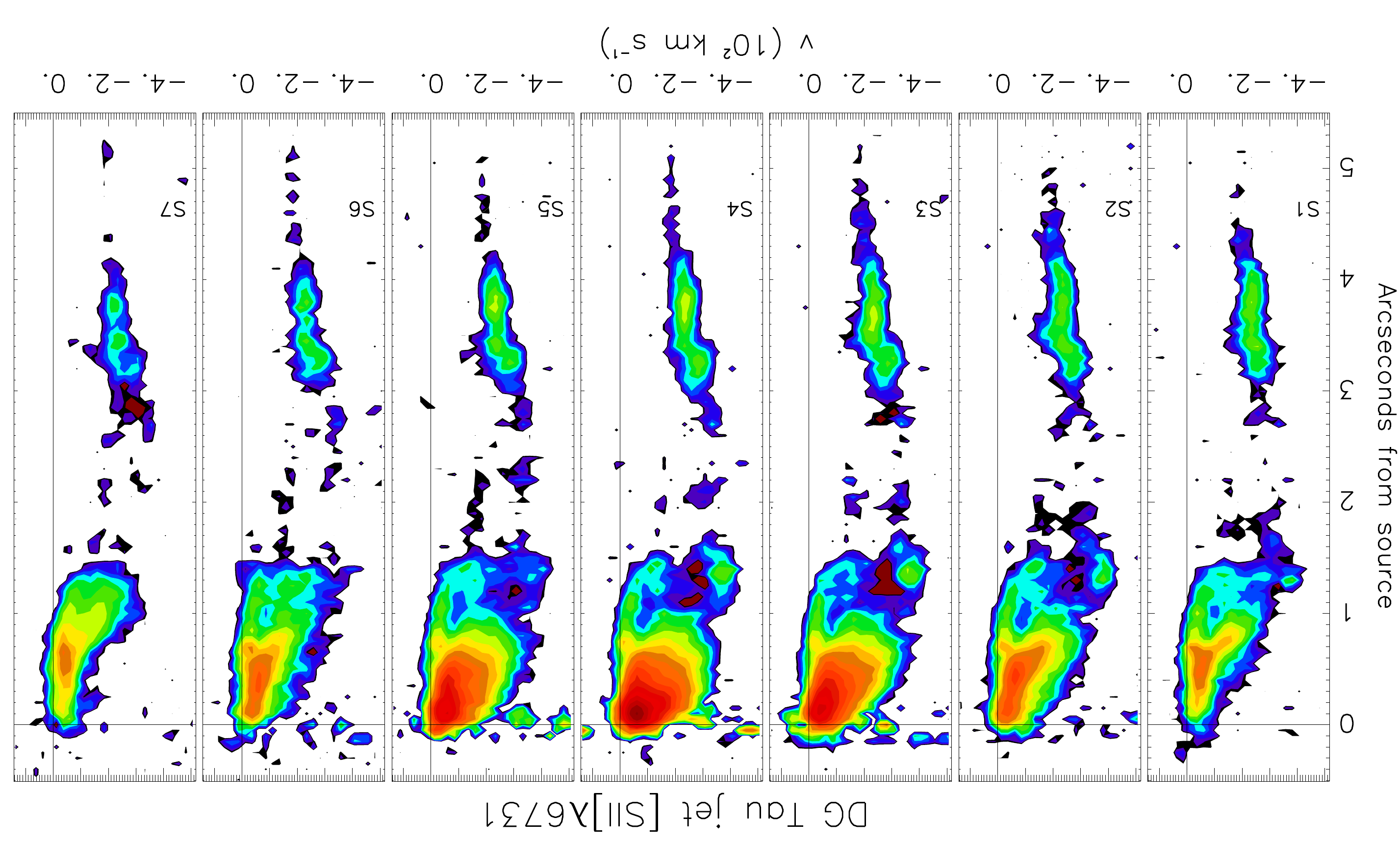}}
  \caption{Continuum-subtracted HST/STIS position-velocity diagrams (PVDs)
of the flow from DG~Tau, in the [\ion{S}{ii}]~$\lambda$6731 line and in
slit positions from S1 to S7 (from southeast to northwest), offset by 0.07 arcsec across the jet width.
The colored contours are from
$1.1\,10^{-15}~\mathrm{erg s}^{-1}\,\mathrm{arcsec}^{-2}\,\mathrm{cm}^{-2}$\,\AA$^{-1}$
(equivalent to $3\sigma$), with a ratio of 2$^{2/5}$. The vertical line
marks zero velocity with respect to the star. The horizontal
lines mark the position of the star and of previously 
identified knots. Adapted from MA14.
}
  \label{DG_Tau_SII}
\end{figure*}

The seven long-slit spectra analyzed in MA14 were taken on January 14 1999, 
keeping the slit parallel to the outflow axis and 
stepping it by $0.7''$ across the jet width.
In this way it was possible to obtain a three-dimensional 
data cube of the optical outflow, with two spatial dimensions 
and one in radial velocity.
The  spectra included several strong forbidden lines, as
[\ion{S}{ii}]~$\lambda\lambda$6731, [\ion{N}{ii}]~$\lambda\lambda$6583, 
[\ion{O}{i}]~$\lambda\lambda$6363, and 
have spectral and spatial  resolution of 0.554 \,\AA\ and 
$\sim 0.1''$, respectively. 
To ease the comparison with synthetic PVDs (see Sect.~\ref{results}),
we report in Fig.~\ref{DG_Tau_SII} the seven observational PVDs for the 
[\ion{S}{ii}]~$\lambda$6731 line. 

The PVDs of the jet show  two distinct regions of high brightness, the first 
one between the source and $1.4''$, called hereafter the first \emph{blob}, 
and the second one between $\sim 3''$ and $\sim 4.5''$, defined hereafter as 
the second blob.
The plasma has  a wide range of velocities (up to $\sim -300$ 
km~s$^{-1}$) close to the source, confirming that most 
of the accelerating process is confined in the very first AUs; 
in the outer slit positions, however,  lower and lower 
velocities are seen, supporting an onion-like kinematic structure  
\citep{Bacciotti00}.
For all the  forbidden lines, and for all the slit positions, 
the surface brightness peaks  at  low velocities at the beginning of the jet.
Further away, the first blob appears to have 
at least two different velocity components, one at constant low velocity, 
and one of progressively higher speed moving away from the star. 
This could be due to either a persisting acceleration or to the fact that progressively 
slower material is emitted from the source as time passes.
The region of low emission 
between $1.4''$ and $\sim 3.1''$ from the star turns out to be   
occupied by tenuous plasma at high ionization and temperature \citep{Dougados00}. 
The second blobshines between   
$\sim 3''$ and $\sim 4.5''$ in all the emission lines, and the 
emission properties suggest that the gas is reheated by shocks (MA14). 
This is also probably the nature of a 
faint third blob observed at $7''$ in the HST spectra,
arising from a previous  emission episode and not analyzed in MA14.

Similar data are available for the bipolar jet from the T-Tauri star RW-Aur,
located in the same star forming region as DG-Tau (at a distance of 140 parsecs). 
The spectra,  taken by HST/STIS with  the same slit setting and processed 
with the same technique used in MA14 to generate the PVDs of the jet from DG-Tau, 
are discussed in \citet{Woitas02,Woitas05,Melnikov09}.
In particular, \citet{Melnikov09} (Fig. 1) provide a set of seven parallel PVDs for 
the [\ion{S}{ii}]\,$\lambda$6731 lines, extending over $4''$ on both sides of the 
source. 

The PVDs show that the jet is highly collimated, and presents a
chain of emitting knots, over the whole jet length, that has been 
widely investigated in the last decade. There is a general agreement about the 
existence of pulsating mechanisms at work at the base of the jet, though some
uncertainty holds about the temporal gap between knots. 
Moving knots in the [SII] lines, down to 56 AU from the source, have been 
observed by \citet{Dougados00}. These knots might be due to temporal fluctuations 
with period ranging from $\sim 5$ to $\sim 10$~yr, \citep{Melnikov09}. 
\citet{Lopez03} have suggested that the inner knot spacing might 
originate by superposition of short timescale random perturbations 
(with period of order 3-10 years), and more regular fluctuations on 
longer timescales of $\approx$ 20 yr. Finally, the correlation between 
time variability outflows in T-Tauri stars, and the driving source 
of unsteady mass accretion has more recently been investigated by \citet{Chou13}.

The jet is highly asymmetric in brightness, velocity, and physical properties. 
The radial velocity peaks at $-180\,\mathrm{km~s}^{-1}$ in the blue lobe and at 
$+100\,\mathrm{km~s}^{-1}$ in the red lobe, and there is not a  spread 
in velocity as wide as in DG-Tau. 
In general, the properties of the RW-Aur jet appear to be very different from those 
in the DG-Tau jet, and we have thus chosen to consider this dataset as a useful test 
of the predictive properties of our numerical procedure. 



\section{Numerical simulations setup} \label{setup}
%
%
\subsection{Equations and algorithms} \label{settings}

We perform numerical simulations of a magnetized, rotating jet 
in cylindrical coordinates $r$ and $z$, 
under the assumption of axisymmetry ($\partial/\partial\phi\equiv 0$). 

With the exception of the test case described in Sect.~\ref{Test_case}, in 
all our simulations the jet propagates into a computational domain of 
$300 \times 1200$~AU in $r-z$, described by a
grid of $420\times 1500$ cells clustered in the nozzle area, with grid density 
that decreases for increasing $z$ and $r$. 
 
The adopted code is PLUTO  \citep{Mignone09}.
PLUTO provides a multi-physics, multi-algorithm modular environment, 
especially tailored for simulations of time-dependent, shocked flows 
in Newtonian or relativistic regime.
The code exploits a general framework designed to integrate a system of 
fluid equations written in conservative form, and based on Godunov-like, 
shock-capturing schemes. The equations can be written as
\begin{equation}
\frac{\partial{\mathbf{U}}}{\partial{t}} = -\nabla \cdot 
\mathbf{T}(\mathbf{U}) + \mathbf{S}(\mathbf{U}),
\label{consequa}
\end{equation} 
where $\mathbf{U}$ denotes the state vector of conservative variables, 
$\mathbf{T(U)}$ is a tensor whose rows are the fluxes of 
each component of $\mathbf{U}$, and $\mathbf{S(U)}$ is the source terms vector.
When using the ideal MHD module, the former system reads 
\begin{equation}
 \mathbf{U} =
\left[ \begin{array}{c}
        \rho\\
	\bf{m}\\
	 \bf{B}\\
	E
       \end{array} \right],
\qquad
 \mathbf{T} =
\left[ \begin{array}{c}
        \mathbf{m}\\
	\mathbf{mv}-\mathbf{BB}+p_t\mathbf{I}\\
	 \bf{vB}-\bf{Bv}\\
	(E+p_t)\mathbf{v}-(\mathbf{v}\cdot\mathbf{B})\mathbf{B}
       \end{array} \right],
\end{equation}
where $\rho$ is the mass density, $\mathbf{v}$ the fluid velocity, $\mathbf{B}$ 
the magnetic field, $\mathbf{m}=\rho\mathbf{v}$ the momentum density, 
$\mathbf{B}$ the unit tensor,
$E=\rho |{\mathbf{v}|^2}/2 + p/(\gamma-1) + |{\mathbf{B}|^2}/2$
the total energy density (assuming an ideal gas law with adiabatic 
index $\gamma$), and $p$ and $p_t = p + |{\mathbf{B}|^2}/2$ are 
the thermal and total pressures, respectively.
The source just contains the geometrical terms appropriate 
to cylindrical coordinates:
since the jet nozzle is placed at a distance from the star where 
gravity effects are negligible, gravity and other body forces 
have been neglected.

The transport equation for the electron density 
\begin{equation}
\frac{\partial N_{e}}{\partial t}+\nabla \cdot (N_e{\mathbf v})={\cal N}
_{ion}{\cal -N}_{rec},
\end{equation}
is also included in the conservative system, allowing us to evolve the 
ionization fraction for the atomic hydrogen specie. The terms 
${\cal N}_{ion}, {\cal N}_{rec}$ represent the ionization and recombination 
source terms.
The radiative cooling processes are described by the \emph{one-ion} 
nonequilibrium model (option "SNeq" into the PLUTO code), which
introduces a simplified radiative cooling source term 
in the energy equation. Such a choice relies on the assumption 
that emissivity, in the region of interest, arises from weakly shocked gas, 
whose temperature should not exceed 75000 K, the maximum value allowed 
in this model. 
The reliability of former assumptions has been checked in the 
test cases TEST$_{DG3}$ and TEST$_{DG4}$ (see Sect.~\ref{Test_case}). 


The system of equations should be closed by the 
solenoidal condition of the magnetic field, $\nabla\cdot\mathbf{B}=0$.
However, in our simulations $B_r=B_z=0$ will be always zero, so that,
thanks to the axisymmetric assumption,
the solenoidal condition is automatically satisfied everywhere and anytime.

The adopted Riemann solver is a simple but robust two-wave solver
(HLL); we used a linear upwind method for reconstruction of primitive variables
(namely $\rho$, $\mathbf{v}$, $\mathbf{B}$, $p$, and $n_\mathrm{H}$)
at cell interfaces, and we adopted the second order accurate Runge-Kutta scheme 
to update the discretized equations in time.

As far as boundary conditions are concerned,
the jet and the outflow as arising from the
extrapolation of the P06 model are injected in our
simulations from the lefthand side ($z=0$) boundary at all 
times, as will be specified further on in Sect.~\ref{profiles} and \ref{tails}.
Along the $z$-axis ($r=0$) we assume standard conditions for
axisymmetry, whereas zeroth-order outflow conditions
are imposed at the outer boundaries in both $r$ and $z$
directions.

\subsection{Summary of disk-wind theory and P06 results}
\label{models}

Numerous theoretical and numerical works demonstrate that
magneto-centrifugal winds can efficiently extract gravitational energy 
and angular momentum from the accreting disk 
\citep[see, e.g.,] []{Blandford-Payne82, Ferreira97, Shu00, Romanova02,
Pudritz07}.
Observations suggest the existence of jets with different
degree of collimation \citep{Dougados02,Dougados04}, in agreement with 
MHD models that predict the collimation of jets depends on the  
initial magnetic configurations.
According to the most popular models, magnetic field lines either 
originate on the star surface and connect with the disk 
\citep[e.g.,] []{Goodson97}, or they only thread the disk \citep{Ouyed97}. 
In the latter case, a variety of radial magnetic configurations are 
possible, according to the power law
\begin{equation}
B_p(r,0)\propto r^{\,\mu-1},
\label{initialfield}
\end{equation}
which describes the behavior of the poloidal magnetic field at the 
surface of the rotating disk. Both magnetic configurations strongly peaked   
in the inner part of the disk, similar to the X-wind configuration by
\citet{Shu00}, and almost flat magnetic profiles  
\citep{Ouyed97} can be generated by changing the value of $\mu$;
the ``classical'' self-similar disk-wind magnetic geometry of 
\citet{Blandford-Payne82} is the intermediate case for 
$\mu=-0.25$, which yields $B_p\propto r^{-5/4}$.

In P06, Pudritz and collaborators investigate four values of $\mu$,
the parameter that drives both the jet mass load and the opening angle, 
or the  collimation degree. The four cases ($0.0, -0.25, -0.50, -0.75$) 
correspond to poloidal magnetic fields more and more steeply declining with $r$. 
In particular, the author shows that magnetic configurations corresponding 
to $\mu=0.0$ and $-0.25$ lead to highly collimated jets, 
whereas $\mu=-0.5$, $-0.75$ generate wide-angle outflows. 
Using the gas parameters at the surface of the rotating accretion disk
as boundary conditions, the authors let the outflow self-generate,
and provide the axisymmetric solution vector
$\mathcal{S}(r,z) = (\mathbf{v}, \mathbf{B}, n_H)$
at the distance $z=72\,r_i$ from the central object,
$r_i$ being the simulation length scale, defined
as the inner disk radius threaded by the magnetic field lines.

\subsection{Initial and boundary conditions} 
\label{inflow}

The boundary inflow conditions at the inlet left side of the 
computational box, and the initial conditions representing 
the unperturbed ISM at time $t=0$ must be initialized 
prior to running numerical simulations of propagating jets. 
While the ISM parameters can be chosen 
in a given range of ``reasonable'' values (and have
weak influence on the results, see Sect.~\ref{results}), 
the inflow conditions at the nozzle play a crucial role, and 
are derived from P06 solutions according to the afore mentioned 
APM procedure. 
In particular, the radial profiles of the hydrogen numerical density
$n_H$, velocities components ($v_z, v_r, v_\phi$),
toroidal magnetic field $B_{\phi}$,
temperature $T$ and ionization fraction $x_e$ 
must be assigned at the position $z=z_\mathrm{nozzle}$.

In P06 the radial profiles of the solution $\mathcal{S}(r,z)$ are 
given in nondimensional units, as functions 
of length, density, and velocity scales, $r_i$, 
$n_H^\mathrm{disk}(r_i)$ and $v_k^\mathrm{disk}(r_i)$, defined, respectively, 
as the disk radius threaded by the inner magnetic field lines 
on the disk surface, and the corresponding gas density and 
Keplerian velocity.
Therefore, once we chose the value of the model parameter $\mu$, and the 
corresponding numerical solution $\mathcal{S}(r,z)$, 
the scale factors must be computed, taking both the physical 
properties of the observed stellar jet and the prescriptions 
given in \citep{Ouyed97} into account.

\subsubsection{The parameter $\mu$ } \label{mu}

The simulations in P06 are driven by the choice of $\mu$, which yields the 
radial profile of the poloidal magnetic field $B_p(r)$ on the disk surface, 
and the outflow collimation degree.
In particular, $\mu=0.0$ (i.e., $B_p\propto r^{-1}$) and $\mu = -0.25$
(i.e., $B_p\propto r^{-5/4}$) generate profiles that
smoothly decrease with $r$, and enforce the collimation effects 
(small angle jets) with respect to cases $\mu=-0.5$ 
(i.e., $B_p\propto r^{-3/2}$) and $\mu = -0.75$ 
(i.e., $B_p\propto r^{-7/4}$), where $B_p(r)$ is peaked in the inner disk 
region and rapidly vanishes when moving outward (wide-angle jets).
Our simulations have considered the four values of $\mu$. Nevertheless, 
since parameters $\mu=0.0$ and $\mu = -0.25$ provided quite similar solutions, 
only results for the case $\mu=-0.25$ have been shown.

\subsubsection{The scale factors} \label{scales}

The scale factors depend on the properties of the jet under investigation, and 
drive the choice of the simulation parameters. Namely, the length scale 
$r_i$ directly affects the nozzle position with respect to the star, 
$z_\mathrm{nozzle}$. In P06 the solution $\mathcal{S}(r,z)$ 
is provided at a distance $\hat{z}=72\,r_i$, beyond which we assume 
that the gross features of the accelerated flow stay unchanged. 
Since the estimated length scale $r_i$ is 0.07~AU for both DG-Tau and RW-Aur, 
we have $\hat{z}=5.04$~AU. As a consequence, the nozzle is supposed to be 
at $z_\mathrm{nozzle} \ge 5.04$~AU from the star. 

The choice of $z_\mathrm{nozzle}$ affects the value of some main inflow
parameters, such as temperature, density, and ionization fraction. 
Their mean value can be computed by using the so-called \emph{BE technique},
which allows us to estimate ionization fraction, electron density and,
hence, total hydrogen density, from the ratios of some optical 
forbidden emission lines \citep[see][]{BacciottiEisl99, Podio06}.
In the recent past, the BE technique has been widely used to obtain
physical parameters of observed jets \citep[see, e.g.,][]{Bacciotti02a,
Hartigan07,Coffey08,Melnikov09}.
In \citet{Bacciotti02a}, the authors find a temperature of DG Tau jet,
in the high-velocity channel and at z=0.3 arcsec from the star,
of $\approx 10^4$~K, with positive gradient toward the source.
In this frame we have decided to assume a value of $1.5\,10^4$~K 
at the nozzle position, for $z=0.2$~arcsec.
Moreover, Fig.~12 and Fig. 13 (Sect 4) of MA14 provide the 
longitudinal profiles of the mean electron fraction and hydrogen density
in the high-velocity channel. 
According to these figures, at $z=0.2$~arcsec we have $x_\mathrm{e}\approx 0.3$, 
and the hydrogen density is $\approx 8\,10^5$~cm$^{-3}$ 
(\citet{Melnikov09} provides a value of 
$\approx 3.2\,10^4\,\mathrm{cm}^{-3}$ for RW-Aur).

The temperature and ionization fraction might be due to the presence 
of a photoionizing X-ray source, \citep{Tesileanu12}. 
The assumption that a photoionizing X-ray flux
is at work is also supported by recent observations of DG-Tau microjet,
which shows a rich X-ray phenomenology with a hard component centered
on the central object \citep{Guedel11}).
The values mentioned earlier in this paper are averaged 
through the body of the jet, and are typical of the high-velocity component. 
We can use them as local values if we assume that, at the nozzle,
the flow is dominated by the high-velocity component. 
These physical parameters are needed to compute velocity, density,
and magnetic scales, and to move from the nondimensional plots of 
P06 to dimensional plots, as explained here in the following.

Concerning the velocity scale, since  
the mass and the length scales are the same for DG-Tau and RW-Aur jets, 
($M_\star = 1 M_\odot$, $r_i = 0.07$~AU),
the velocity scale defined as the Keplerian velocity on the disk surface 
at radius $r_i$ for both jets is 

\begin{equation}
v_k^\mathrm{disk}(r_i)=\sqrt{GM_\star/r_i} = 112\,\mathrm{km~s}^{-1}.
\end{equation}

The density scale $n_\mathrm{H}^\mathrm{disk}(r_i)$ used in P06 
cannot be provided by direct observations of the disk surface, 
instead, it can be inferred by matching some reference point in the 
nondimensional radial density profiles from P06 with real data. 
So, if we make reference to the density bulk close to the axis
($\approx 0.0013$ for all values of $\mu$, see P06, Fig.~2b),
the scale factors are obtained by dividing the above cited 
values by 0.0013, leading to 
$n_\mathrm{H}^\mathrm{disk}(r_i) = 6.2\,10^8$~cm$^{-3}$
for DG-Tau, and $n_\mathrm{H}^\mathrm{disk}(r_i) = 2.4\,10^7$~cm$^{-3}$ 
for RW-Aur. 
It stands to reason that these numbers suffer from some uncertainty,
though results from numerical simulations have shown to be 
robust with respect to small changes in their values. 

Once $n_\mathrm{H}^\mathrm{disk}(r_i)$ and $v_k^\mathrm{disk}(r_i)$ 
are computed, the magnetic field scale, defined as the magnetic 
field strength on the disk surface at radius $r_i$, 
can finally be evaluated as \begin{equation}
B^\mathrm{disk}(r_i)= v_k^\mathrm{disk}(r_i) 
\sqrt{\frac{8 \pi \, n_\mathrm{H}^\mathrm{disk}(r_i) 
\,\mu_w m_\mathrm{p}}{\beta\, \delta}}
\end{equation}
where $\mu_w$, the mean molecular weight, depends on the relative
metal abundance in the accreting disk ($\mu_w = 1.4$  
for solar-like abundances),
$m_\mathrm{p}$ is the proton mass, $\beta$ is the  
gas to magnetic pressure ratio, and $\delta$ is the Keplerian kinematic
to thermal energy density ratio, taken 
at the distance $r_i$ from the central object on the disk surface. 
Assuming $\beta=1/3$ and $\delta=300$, as in P06, we find 
$ B_i^\mathrm{disk}(r_i)=0.18$~G for DG-Tau and 0.04~G for RW-Aur.

The set up parameters and the normalization factors are listed in 
Tab.\,\ref{inflowparam}.

\begin{table}[!htb]
\begin{center}
\begin{tabular}{|c|c|}
 \hline
  ISM density & 2.5 10$^4$ cm$^{-3}$ \\
  \hline
  ISM  temperature & 5 10$^2$ K \\
  \hline 
  ISM  ionization fraction& 0.0\\
  \hline
  DG-Tau and RW-Aur central mass & 1.0 $M_{\odot}$ \\
  \hline
  DG-Tau and RW-Aur length scale  & 0.07 AU \\
  \hline
  DG-Tau and RW-Aur velocity scale & 112 km s$^{-1}$ \\
  \hline
  DG-Tau density scale& $6.2\,10^8$ cm$^{-3}$\\
  \hline
  RW-Aur density scale& $2.4\,10^7$ cm$^{-3}$ \\
  \hline
  DG-Tau magnetic field scale& 0.18 G \\
  \hline
  RW-Aur magnetic field scale& 0.04 G \\
  \hline
 \end{tabular}
\end{center}
\caption{Initialization parameters and scale factors.}
\label{inflowparam}
\end{table}

\subsubsection{The inflow radial profiles} \label{profiles}

\begin{figure*}[tbp]
\centering
\hspace{-0.5cm}
\subfigure{}\includegraphics[width=7.5cm]{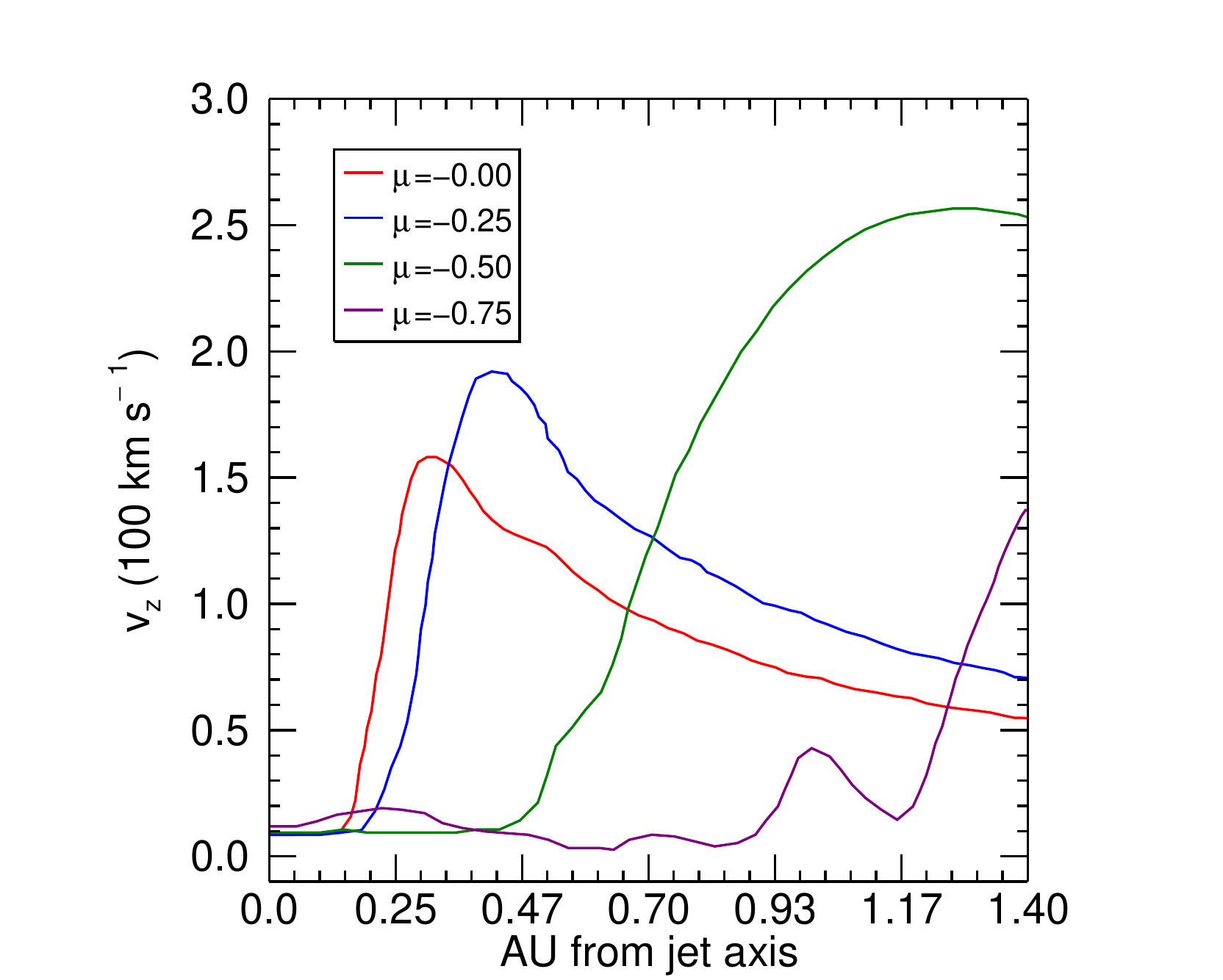}\quad%
\hspace{-1.3cm}
\subfigure{}\includegraphics[width=7.5cm]{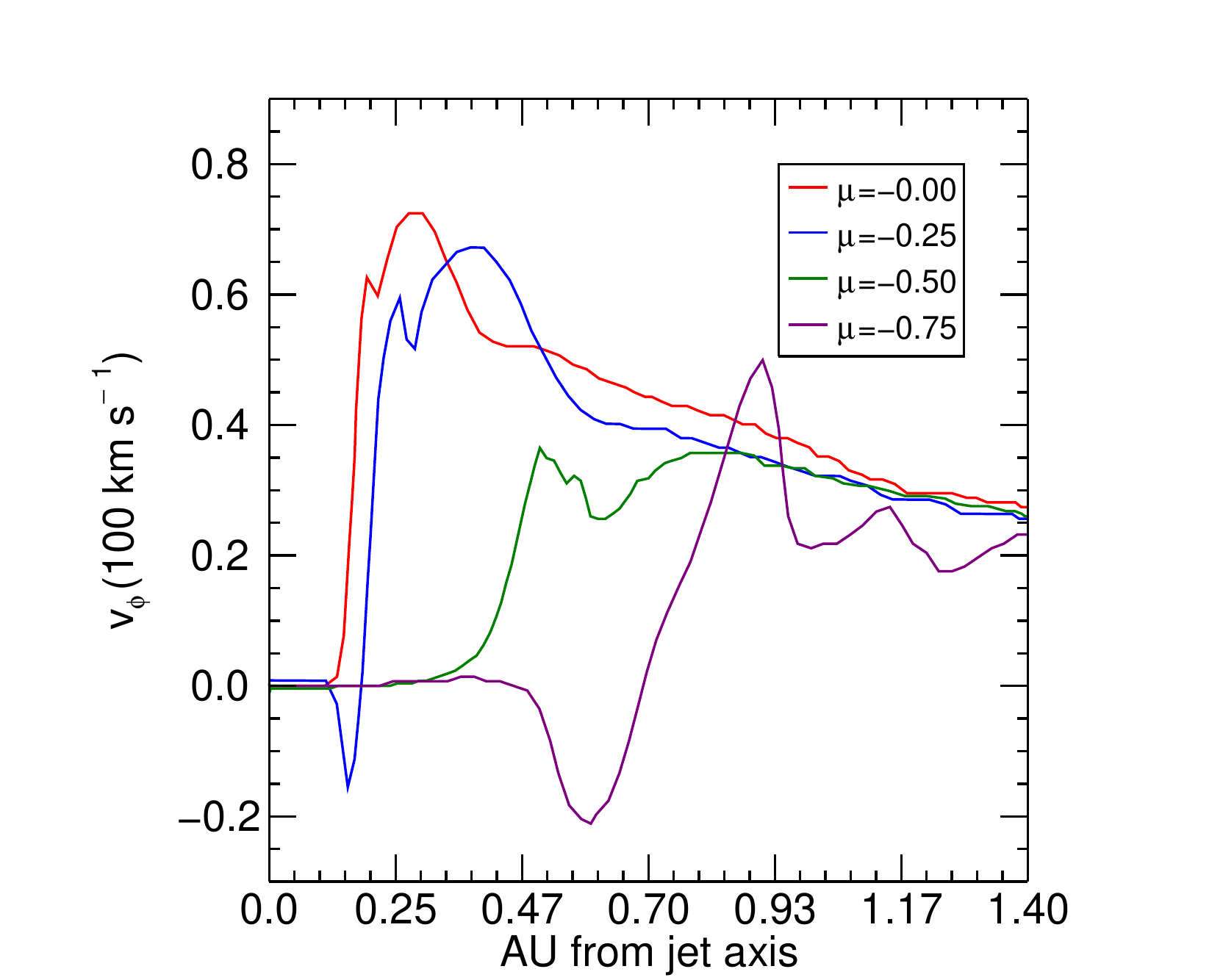}\\%
\hspace{-0.5cm}
\subfigure{}\includegraphics[width=7.5cm]{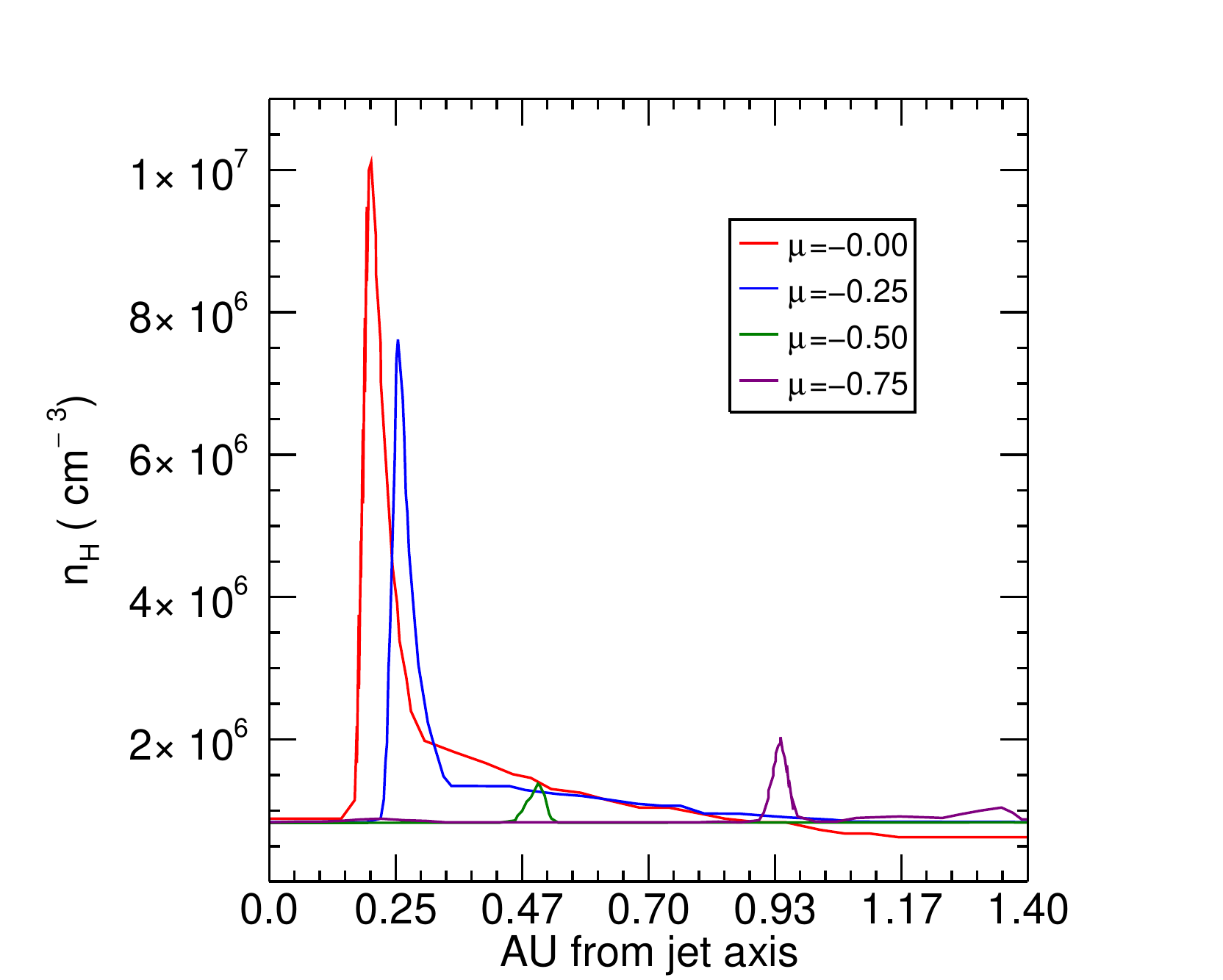}\quad
\hspace{-1.3cm}
\subfigure{}\includegraphics[width=7.5cm]{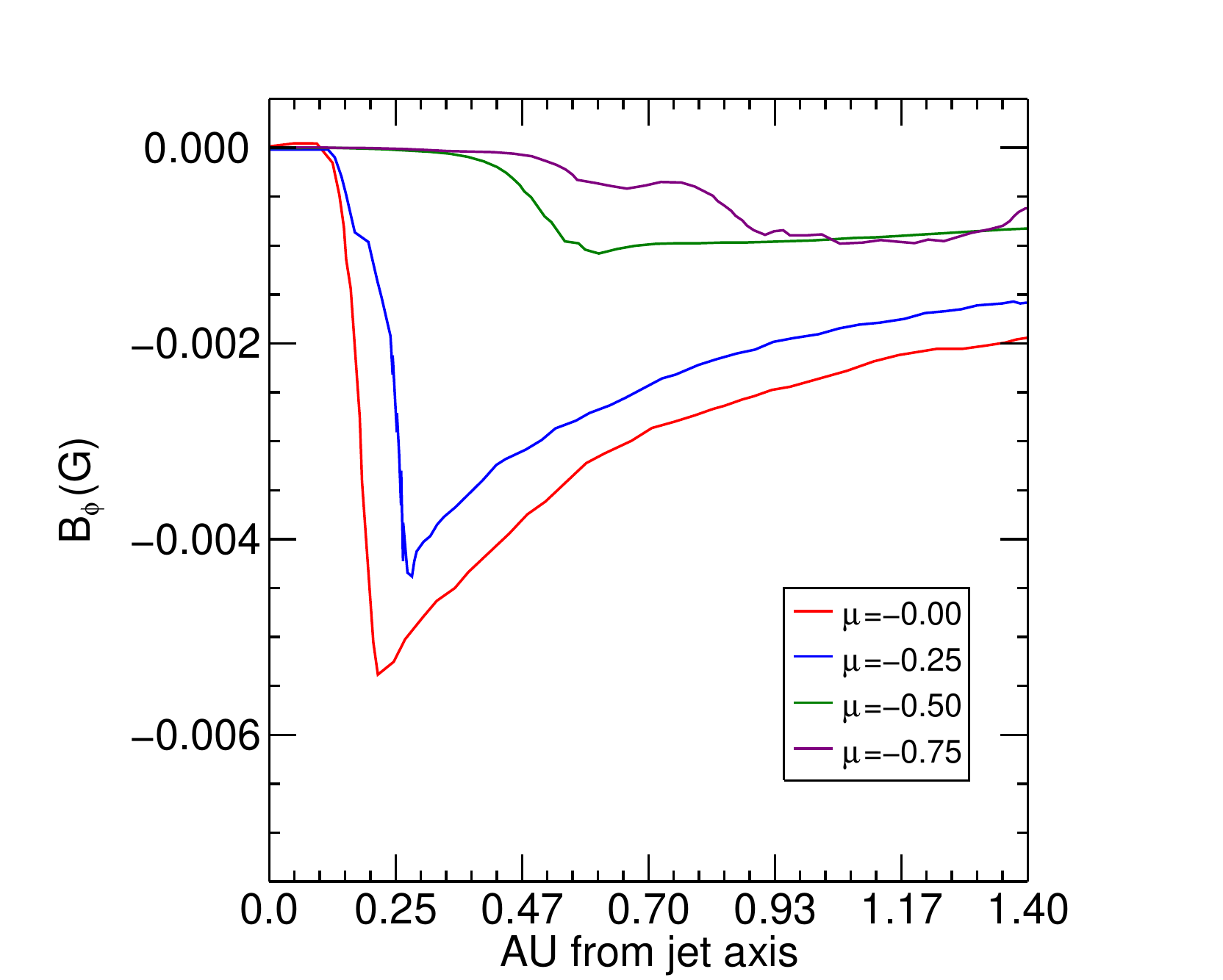}%
\caption{Dimensional radial profiles at the nozzle from P06,
for each value of the parameter $\mu$.}
\label{inflow_profiles}
\end{figure*}

After computing the scale factors, the solution vector arising from 
the P06 model 
$\mathcal{S}(r,z_\mathrm{nozzle})=(\mathbf{v},\mathbf{B},n_\mathrm{H})$, 
is retrieved, 
and the dimensional radial profiles of 
$n_H(r,z_{nozzle})$, $v_z(r,z_{nozzle})$,
$v_{\phi}(r,z_{nozzle})$ and $B_{\phi}(r,z_{nozzle})$
can be injected into the simulation box. In all our simulations, we set 
$v_r(r,z_{nozzle})$, $B_z(r,z_{nozzle})$, and $B_r(r,z_{nozzle})$ 
equal to zero. In fact, the nozzle is far enough from the source 
to assume that a purely toroidal magnetic field is generated from the 
poloidal configuration on the disk \citep[see, e.g.,][]{Pudritz06, Zanni07}, 
and that the flow is aligned with the axis.
Fig.\,\ref{inflow_profiles} shows the dimensional inflow
profiles of the solution vector $\mathcal{S}(r,z_\mathrm{nozzle})$
in the inner region described in P06, up 1.4 AU from the axis.

We used the same shape for both
the temperature $T(r,z_\mathrm{nozzle})$ 
and the ionization fraction $x_e(r,z_\mathrm{nozzle})$ profiles,
with constant values within the first 2 AU from the axis 
(15000 K and 0.4, respectively) and linear extrapolation up to 
$r\approx 67$~AU, where the gas parameters match the values of the ISM 
and stay unchanged up the border of the numerical box, which is located
300 AU far from the axis (see the discussion in Sect.\,\ref{tails}).

\subsubsection{The ``tails''} \label{tails}
As shown in Fig.\,\ref{inflow_profiles}, the inflow profiles 
do not go beyond the point $r = 20 r_i$
(1.4 AU for both DG-Tau and RW-Aur), while the 
numerical box is 300 AU large. 
Therefore, one more degree of freedom arises from the choice of 
the function used to match inner and outer regions (from here on, 
the ``tail''). Tails are meant to represent the wind lifted
from the disk.
Coaxial low velocity, wider winds seen in [\ion{H}{2}] lines are observed in many stellar 
outflows in the IR spectral range
\citep[see, e.g.,][]{Takami04,Agra-Amboage11}. 
As an example, \citet{Takami04} reports a molecular outflow 
from DG-Tau thermalized at $2000~K$ that extends 
up to $\sim 50$~AU from the jet axis in the transverse direction.
A detailed study of low-velocity winds, together with their
effects on the emission properties, is beyond the aim of this work. 
Here, we only show that tails play a crucial role to reproduce 
the observed jet features.

We use two different kinds of tails for our simulations:
\begin{itemize}
\item[-] Linear tail: profiles in the inner region decrease linearly from 
1.4 AU up to 67 AU (1 code unit), where they match the ISM values. Beyond
this point they stay constant up to the numerical box domain, 300 AU 
far from the axis.
\item[-] Exponential tail:  same as linear tails, with exponential decrease 
from 1.4 to 67 AU.
\end{itemize} 
\subsection{The post-processing code OTS} \label{ppcode}

The PLUTO code provides, cell-by-cell, the physical parameters in the 2-D 
numerical domain. The post-processing code OTS
inputs the values of the physical parameters in the gas 
and computes the emission producing synthetic
PVDs of forbidden emission lines.

This goal is achieved in two steps. 
In the first step, we calculate the luminosity cell by cell in three
selected forbidden lines commonly observed in these objects, that is 
[\ion{S}{ii}]$\lambda6731$, [\ion{N}{ii}]$\lambda6583$, and
[\ion{O}{i}]$\lambda6363$.
To this aim, we follow the procedure described in 
\citet{BacciottiEisl99}. Briefly,
the ionisation state of oxygen and nitrogen is calculated 
by a dedicated routine that considers charge-exchange with hydrogen, 
collisional ionisation and radiative
and dielectronic recombination. Charge-exchange is the dominant mechanism
for oxygen, while for nitrogen the contribution of the different
processes is comparable.
For both ions,  the ionisation fraction turns out to be a function of
$T_\mathrm{e}$ e and $x_\mathrm{e}$.
Because of its low ionisation potential, sulphur can be considered totally
ionised once in the regions of interest. 
The emissivities in the selected lines are calculated
determining the electronic level population through the statistical
equilibrium equations
applied to a five-levels atom model. Elemental abundances are taken from
\citet{Osterbrock89}.

In the second step, after rotating the map of the emissivity in the $(r,z)$ plane around the $z$-axis,
to generate the azimuthal dimension $\phi$, emissivities are summed up 
along the line of sight, and synthetic PVDs are created. In the present case, to simulate HST/STIS, 
cell-by-cell integration occurs inside ``coring tubes'' whose section 
mimics the HST angular resolution, $0.1'' \times 0.1''$
($14$~AU$\times 14$~AU 
at the distance of DG-Tau and RW-Aur), is inclined in  
the cylinder with the proper angle for the line of sight 
(38$^{\circ}$ w.r.t. the jet axis for DG-Tau, 46$^{\circ}$ for RW-Aur), and intercepts
the ($r, z$) plane, perpendicular to the slit "plane of sight", 
at a distance $z_c$ from the origin. 

The considered forbidden lines have low radiative transition probabilities
and at the densities  retrieved in our
simulations ($n < 10^{9-10}$~ cm$^{-3})$ the medium can
be considered optically thin for those lines.
The OTS code, however, calculates the population of the levels in statistical
equilibrium, also allowing for a correct determination of the emissivities above
the critical density for collisional de-excitation.
Therefore, once the numerical cells contained in each tube have been 
identified, the corresponding emissivities, multiplied by the cell volumes,
are summed up and distributed into velocity channels. Such a procedure yields the 
emissivity/radial velocity histogram for a given value of $z_c$, or
for a given coring tube in the same slit position. 
Integrated emissivities (in erg~s$^{-1}$) must be 
divided by $4 \pi r^2$ and by the HST pixel extension in arcsec
to obtain the total surface brightness of the numerical jet in the 
proper units, erg~s$^{-1}$~arcsec$^{-1}$~cm$^{-2}$.
Finally, moving $z_c$ along the axis, and stepping the parallel planes 
of view of $0.7''$, leads to the seven different PVDs, labeled 
from S1 to S7 as in MA14.

The $z$ component in synthetic PVDs
represents the tangential dimension, or the $z$ component in the
jet frame projected onto the plane of the sky.
Velocities, as usual, must be interpreted as observed radial velocities,
that is to say, the total velocity in the jet frame projected onto
the line of sight.
The velocity channels range from zero to the largest value 
measured in a given object. Their size depends on the spectral 
resolution of STIS spectra (0.554\,\AA\, corresponding to 
$\sim 25$~km~s$^{-1}$ for the [\ion{S}{ii}]$\lambda 6731$ line).

The structure of the OTS code allows us to compute parallel and 
perpendicular PVDs of different species, as well as 2-D images of 
the surface brightness. 
Moreover, different objects and/or observational instruments can 
be simulated, since both the instrumental parameters 
(spatial and spectral resolution), and the observational features 
(distance, angle of sight) can be changed quite easily.

\section{Results} \label{results}

\subsection{Stationary and time-dependent simulations} \label{unsteady}

As already pointed out, we have focused on the microjets 
of DG-Tau and RW-Aur in the bright region up to 4-5~arcsec from 
the source ($\sim 500$~AU at the distance of 140~pc from Earth). 
For DG-Tau, observed PVDs reveal 
the existence of a first almost stationary emission blob within
the first 2~arcsec from the star. The second moving blob
observed at $\sim 4$~arcsec downward the flow, however,
might be due to temporary  fluctuations of the ejection mechanism. 
The jet from RW-Aur shows a pattern of moving knots too,
that probably reflects fluctuations with a period of a few years.
In this frame we have first performed stationary simulations,
to reproduce the "wiping-out" effects that 
long-period "macrojets" may have on the ISM, and the consequent 
formation of steady structures. Eventually, we superimposed time-dependent 
simulations  to generate the nonstationary observed features.


\begin{table}[!htb]
\begin{center}
\begin{tabular}{|c|ccccc|}
\hline
Run & $\mu$ & Tails & 
$V_\mathrm{nozzle}$& $B_\mathrm{nozzle}$ & Steady I.C. \\
& & &   [km~s$^{-1}$] & [G] & \\ 
\hline
DG1 &  $   0.0  $ & exp &155 &  0.014 &yes \\
         &  $ -0.25$ &  lin  & 190 &  0.011 &        \\
\hline
DG2 &  $-0.50$ & exp & 255 & 0.0025 &yes \\ 
\hline
DG3 &  $-0.50$ & lin & 255 & 0.0025 &yes \\ 
\hline
DG4 &  $-0.50$ & lin & 255 & 0.0025 &no \\ 
\hline
RW1 &  $-0.50$ & lin & 255 & 0.0005 & no\\
\hline
RW2 &  $-0.25$ & lin & 190 & 0.002& no\\
\hline
RW3 &  $0.0$ & lin & 155 & 0.0027& no\\
\hline
RW4 & SC & lin & 130 & 0.0036& no\\
\hline
\end{tabular}
\end{center}
\caption{List of parameters used in the various runs.}
\label{runlist}
\end{table}

Numerical simulations are listed in 
Tab.\,\ref{runlist} with labels that refer to the jets
under investigation, DG for DG-Tau and RW for RW-Aur. The 
parameter $\mu$, the kind of tail, the maximum velocity and 
magnetic field at the nozzle are reported with a flag, in 
the last column, indicating the type of inflow conditions imposed
(steady or time-dependent).
The simulation parameters are summarized in
Tab.\,\ref{inflowparam}, whereas the inflow
conditions are shown in Fig.\,\ref{inflow_profiles}.
In most cases, only the "paradigmatic" 
[\ion{S}{ii}]$\lambda$6731 forbidden line emission has been shown. 
Forbidden lines for the [\ion{O}{i}]$\lambda$6300 
are reported in the most representative 
cases only.

Simulations from DG1 to DG4 are performed by using the 
physical parameters that are typical of the jet from DG-Tau, 
whereas those from RW1 to RW4 refer to RW-Aur data.
Taking $M_\star =  M_\odot$ and $r_i = 0.07$~AU,
for both stellar jets,
makes the velocity scale, and $V_\mathrm{nozzle}$, to depend on $\mu$ only. 
In particular, velocity up to to 250~km~s$^{-1}$ arises from wide-angle accelerating 
models, ($\mu$ = -0.5), whereas collimated outflows ($\mu$ close to 0) 
show lower speed.

\subsection{Simulations of the DG-Tau case: collimated jets} 
\label{mu0}
The case DG1 is representative of four different simulations,
where the models $\mu = 0.0$ and $\mu = -0.25$ are explored 
with both exponential and linear tails. These cases provided quite 
similar results, therefore we gather them using the same label.

\begin{figure}[!htb]

 \resizebox{\hsize}{!}{\includegraphics{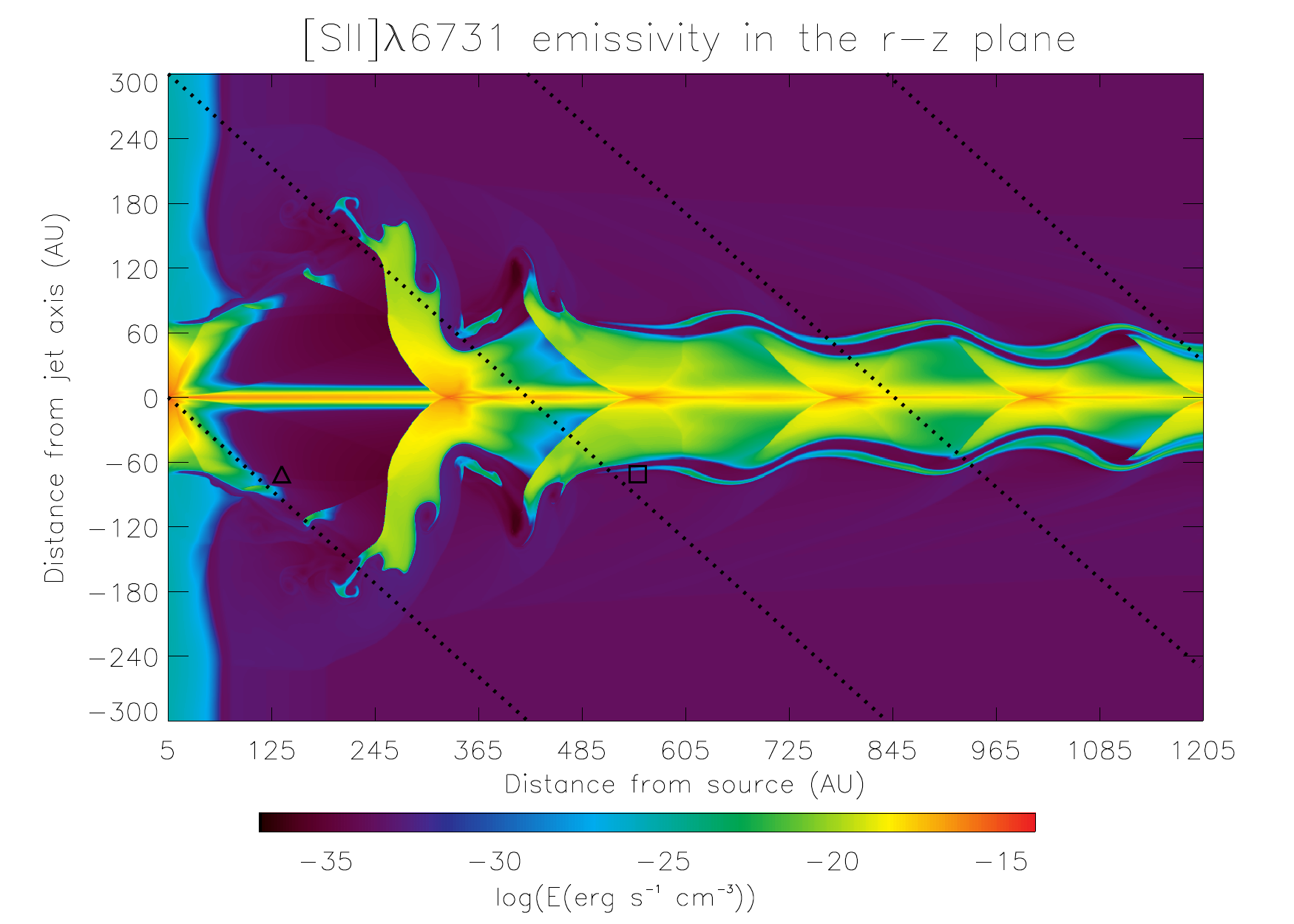}}
        \caption{2-D emissivity in the $\mu=0$ for the
	         exponential tail case. Dotted lines 
		 show the integration path along the line of sight, 
		 whose angle with the jet axis is 38$^{\circ}$.  
		 The lines marked with a triangle and a square 
		 correspond to positions $z=0$ and $\sim 3.4$~arcsec in the PVD. }
	         \label{emissivity_0_exptail}
		    \end{figure}

Fig.\,\ref{emissivity_0_exptail} represents the emissivity (or, the 
numerical emission power density) for [\ion{S}{ii}] $\lambda$6731 in the r-z plane 
for a stationary jet, $\sim 400$ years old. 
Most of the optical emission within the first $\sim$250 AU 
from the source comes from the red-colored, internal beam 
surrounding the axis (hereafter, the \emph{beam}), and from the 
cone-like expansion region close to the nozzle, fueled by the 
injected hot gas (the \emph{cone}). The light-blue region 
close to the left boundary, generated by the pressure and 
density boundary values, is too weak to affect the synthetic PVD. 

Downward from the nozzle, at $z > 250$~AU, internal oblique shocks, 
typical of collimated jets, drive the formation of a pattern of emitting knots 
(in the picture, visible in red) slowly moving with the flow, \citep{Rubini07}, 
while the surrounding green-colored cocoon is filled with weakly emitting gas. 
The outer dark-blue colored, not-emitting part of the domain, represents 
the empty and cold region where the ISM has been stripped off
by the bow-shock.

However, synthetic PVDs look quite different with respect to 2-D emissivity maps
because of the integration along the line of sight (represented with dotted lines)
and the splitting into the velocity channels. The surface brightness contour 
levels, too, are those used in Fig.\,\ref{DG_Tau_SII}. 

\begin{figure}[!htb]
\resizebox{\hsize}{!}{\includegraphics{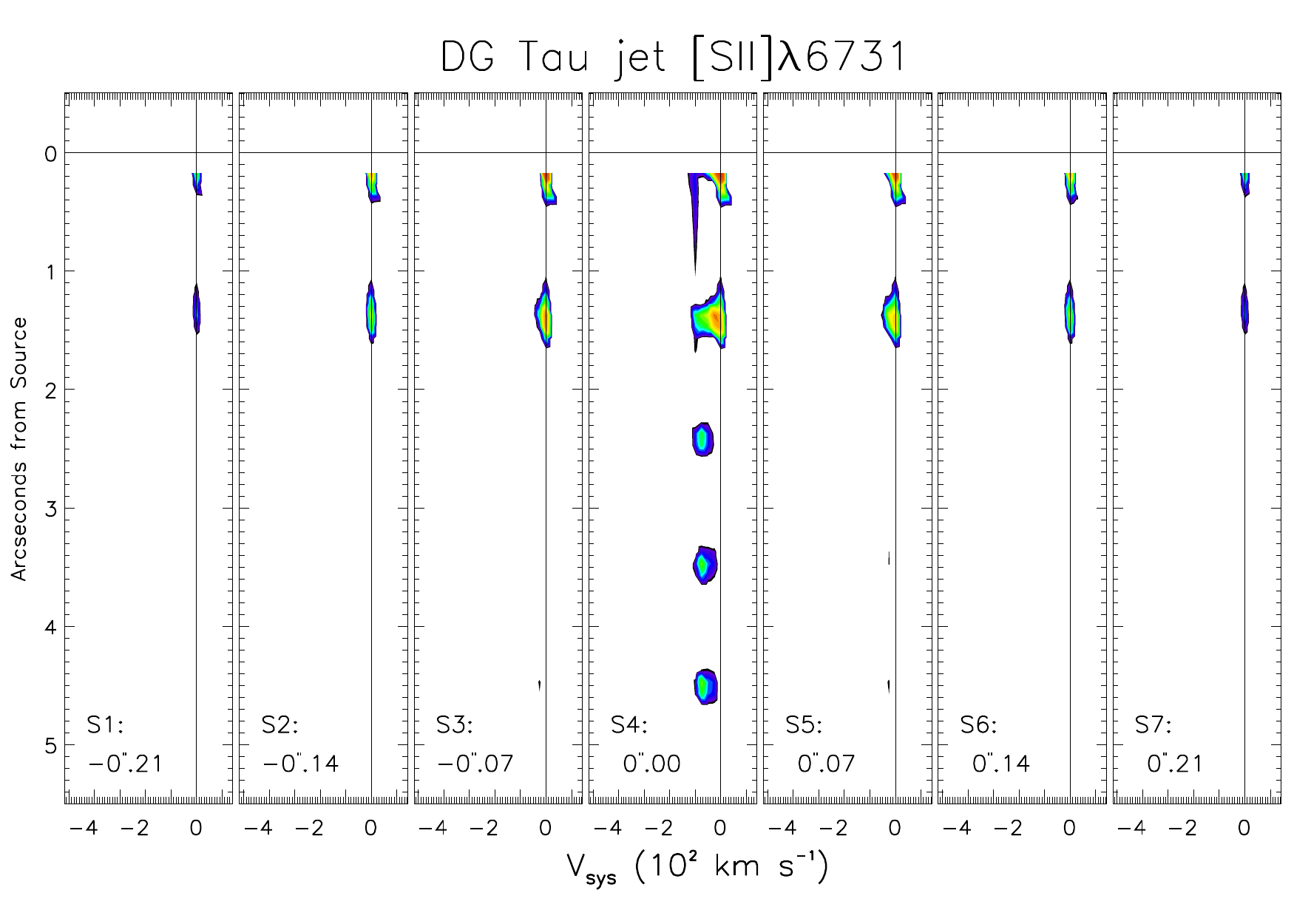}}
\caption{Synthetic PVDs obtained for the $\mu=0.0$ case with
exponential tail added to profiles in P06. 
Contour levels and reference lines are as
in Fig.\,\ref{DG_Tau_SII}, as well as the spatial and spectral resolution. }
\label{pvd_0_exptail}
\end{figure}

Results for DG1 case, shown in Fig.\,\ref{pvd_0_exptail}, can be summarized
as follows:
\begin{itemize}
\item Propagating jets for $\mu=0.0, \,-0.25$ keep a high degree 
of collimation. Internal oblique shocks form, which are not visible 
in lateral slit positions;
\item The emitting zones are too weak and small 
to generate surface brightness comparable with observations, except for 
the central slit position, where a faint chain of internal-shocks-driven 
knots is seen, there is almost no emission;
\item The [\ion{S}{ii}]$\lambda6731$ surface brightness is 
concentrated in the low velocity channel only.
\end{itemize}
In conclusion, the four cases labelled as DG1 do not match the observational
features and, independent of the type of tail, models $\mu=0.0$ and $\mu=-0.25$ 
do not reproduce the properties of the jet from DG-Tau. 

\subsection{Simulations of the DG-TAU case: wide-angle jets}

\begin{figure}[!htb]

 \resizebox{\hsize}{!}{\includegraphics{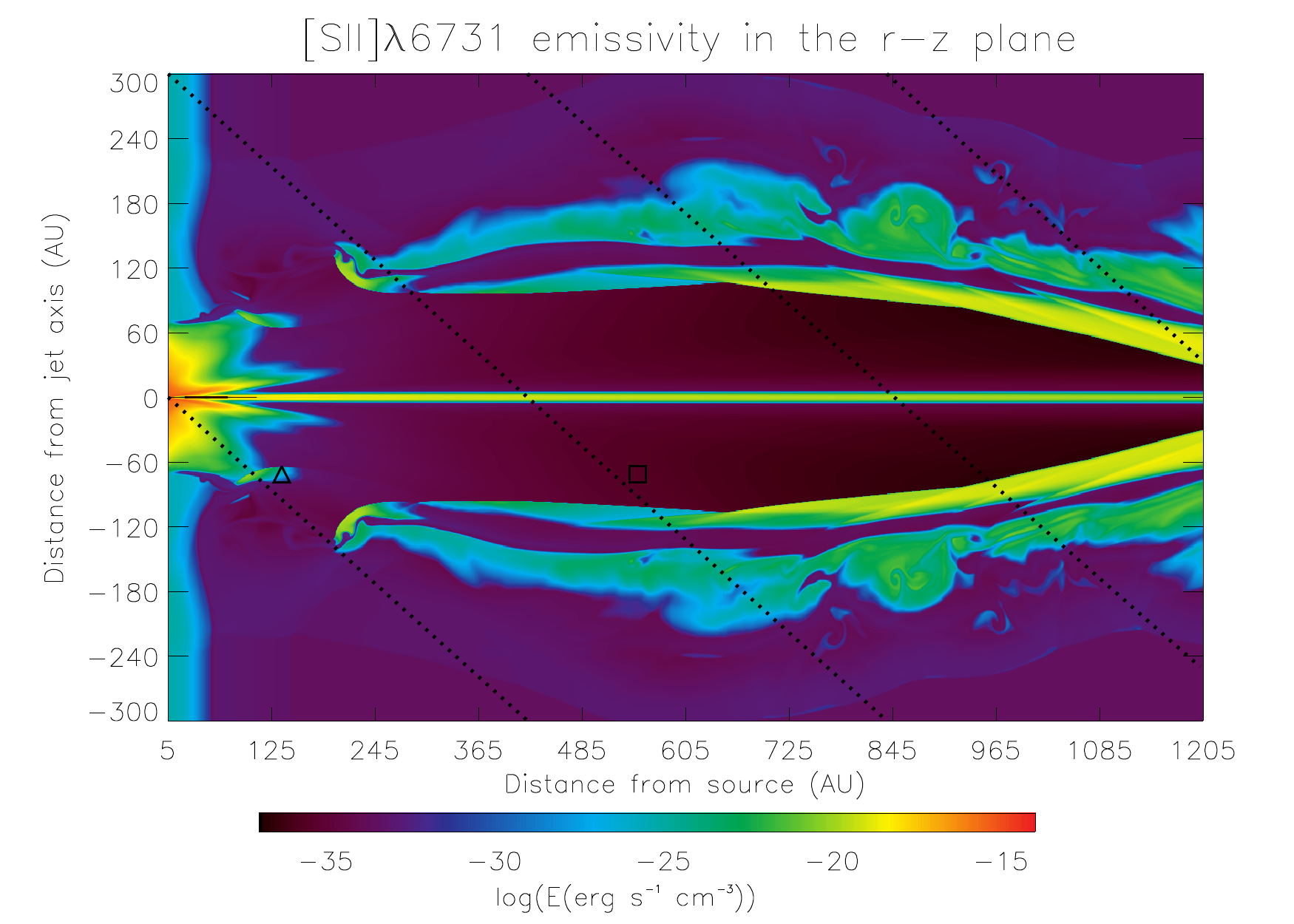}}
       \caption{2-D emissivity map in the $\mu=-0.5$ case. 
Inflow conditions correspond to exponential tails.
Dotted lines show the integration path along the line of sight,
whose angle with the jet axis is 38$^{\circ}$.
Lines marked with a triangle and a square
correspond to positions $z=0$ and $\sim 3.4$~arcsec in the PVD. } 
         \label{emissivity_05_exp}
   \end{figure}

\begin{figure}[!htb]

 \resizebox{\hsize}{!}{\includegraphics{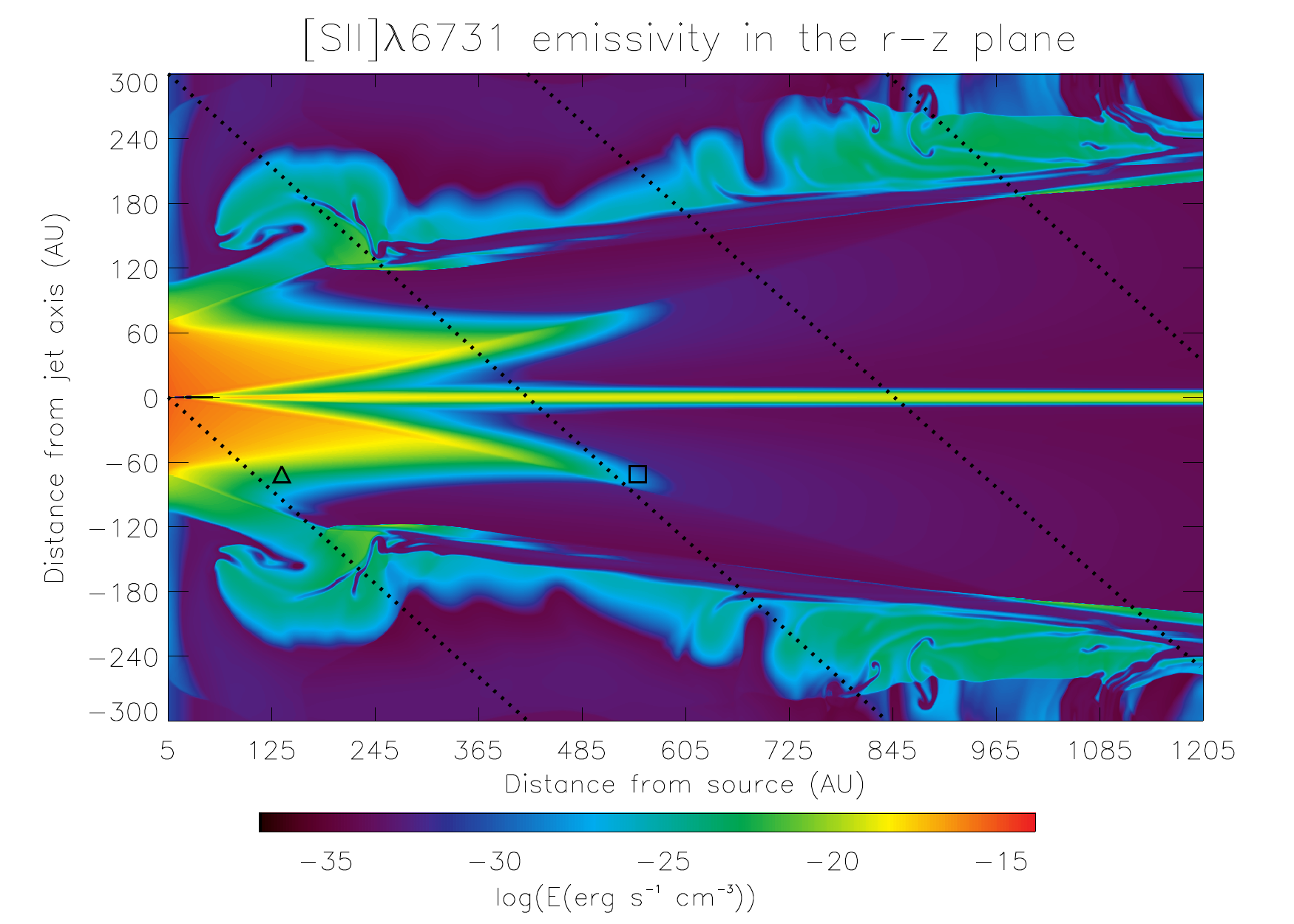}}
       \caption{Same as Fig.~\ref{emissivity_05_exp}, with  
linear tails added to the P06 inflow profiles.}
         \label{emissivity_05_lin}
   \end{figure}

The model $\mu=-0.5$ corresponds to "wide-angle" outflows
(according to the definition given in P06), and is a better candidate to reproduce the real jets
with respect to the $\mu=0$ case, since it is suitable to generate 
faster jets, with velocity in the range of DG-Tau, 
see Tab.\,\ref{runlist} and Fig.\,\ref{inflow_profiles}.
Figs.\,\ref{emissivity_05_exp} and \ref{emissivity_05_lin} 
show the 2-D emissivity for cases DG2 and DG3, corresponding to
exponential and linear tails, respectively.

As in the case of DG1, the surface brightness still originates from 
cone, beam, and cocoon, but in the case of $\mu=-0.5$ no internal shock is present, 
and the less collimated outflow unwraps downward from the nozzle.
This different behavior is measured by the function that we have  
called \emph{collimation function} 
$F_c(r)$, defined as follows: given the cylinder aligned with the axis $z$, 
of radius $r$, and length $\Delta z$ (the longitudinal cell size), 
$F_c(r)$ is the radial flux 
of the radial momentum across the cylindrical surface of area 
$2  \pi  r \Delta z$, 
divided by the longitudinal flux of the 
longitudinal momentum across the circle of area $\pi r^2$.
We thus write
\begin{equation}
F_c(r) = \frac {
            \int _0^{2 \pi} {\rho(r,z,\phi) v_r(r,z,\phi)\, r \,d\phi \,\Delta z }
                  }
                  {
             \int _0^{2 \pi} \int _0^r {\rho(r',z,\phi) v_z(r',z,\phi) 
             \, r' \,dr' d\phi}
                  }.
\end{equation}

\begin{figure}[!htb]
\resizebox{\hsize}{!}{\includegraphics{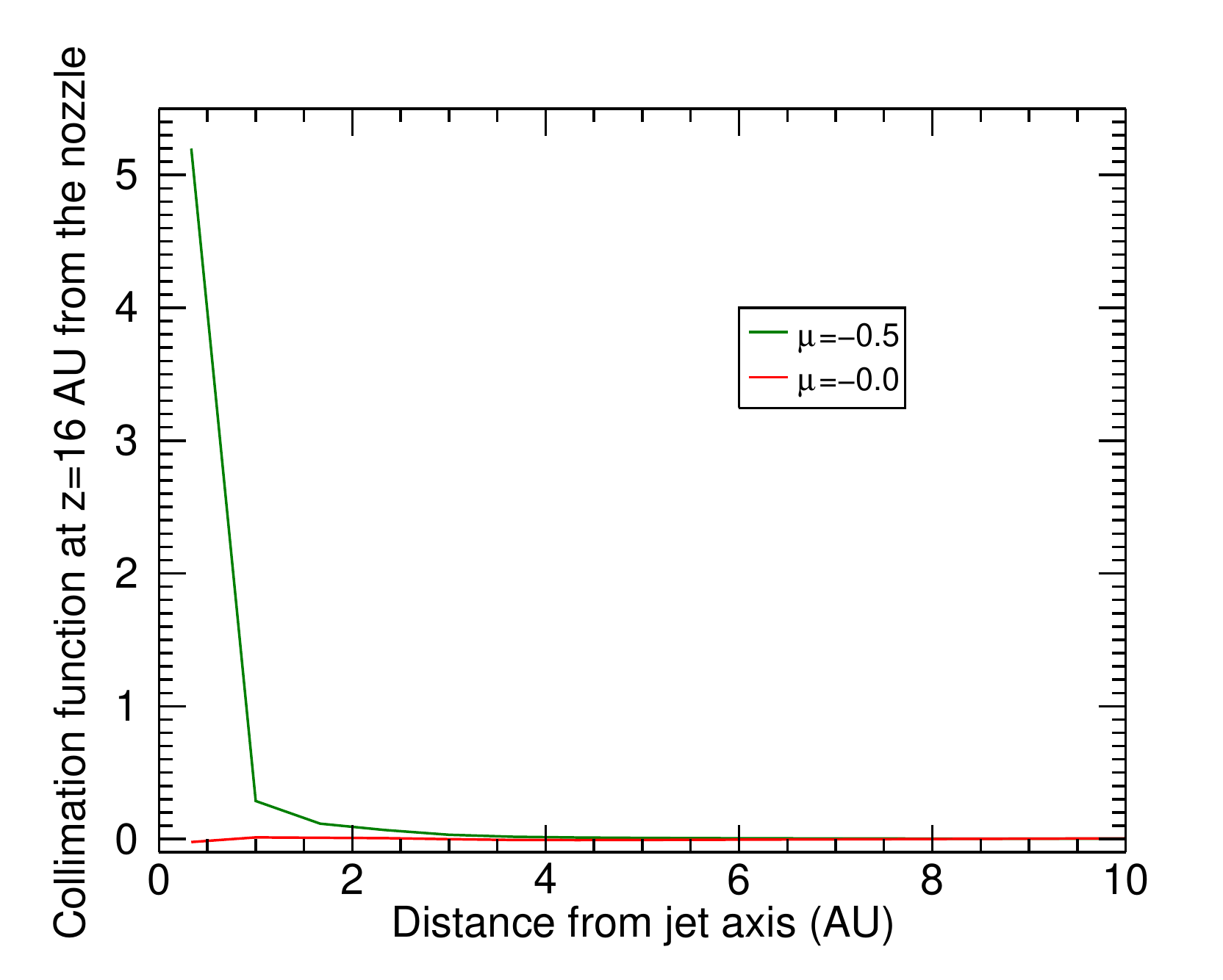}}
\caption{Collimation function {\it vs.} $r$ in $\mu=0.0$ and $-0.5$ cases. The solid line 
shows that in wide-angle jets ($\mu=-0.5$) the gas is pushed outward 
by the total pressure gradient. 
}
\label{coll_profile}
\end{figure}

Fig.\,\ref{coll_profile} shows $F_c(r)$ computed at a distance $z=16$~AU 
from the nozzle. The model $\mu =0$ is, substantially, in 
radial equilibrium, whereas in the $\mu=-0.5$ case the gas is  
pushed outward, the outflow unfolds downward from the nozzle and 
generates the empty cavity surrounding the beam, which is visible in    
Fig.\,\ref{emissivity_05_exp} and Fig.\,\ref{emissivity_05_lin}.

\begin{figure}[!htb]
\resizebox{\hsize}{!}{\includegraphics{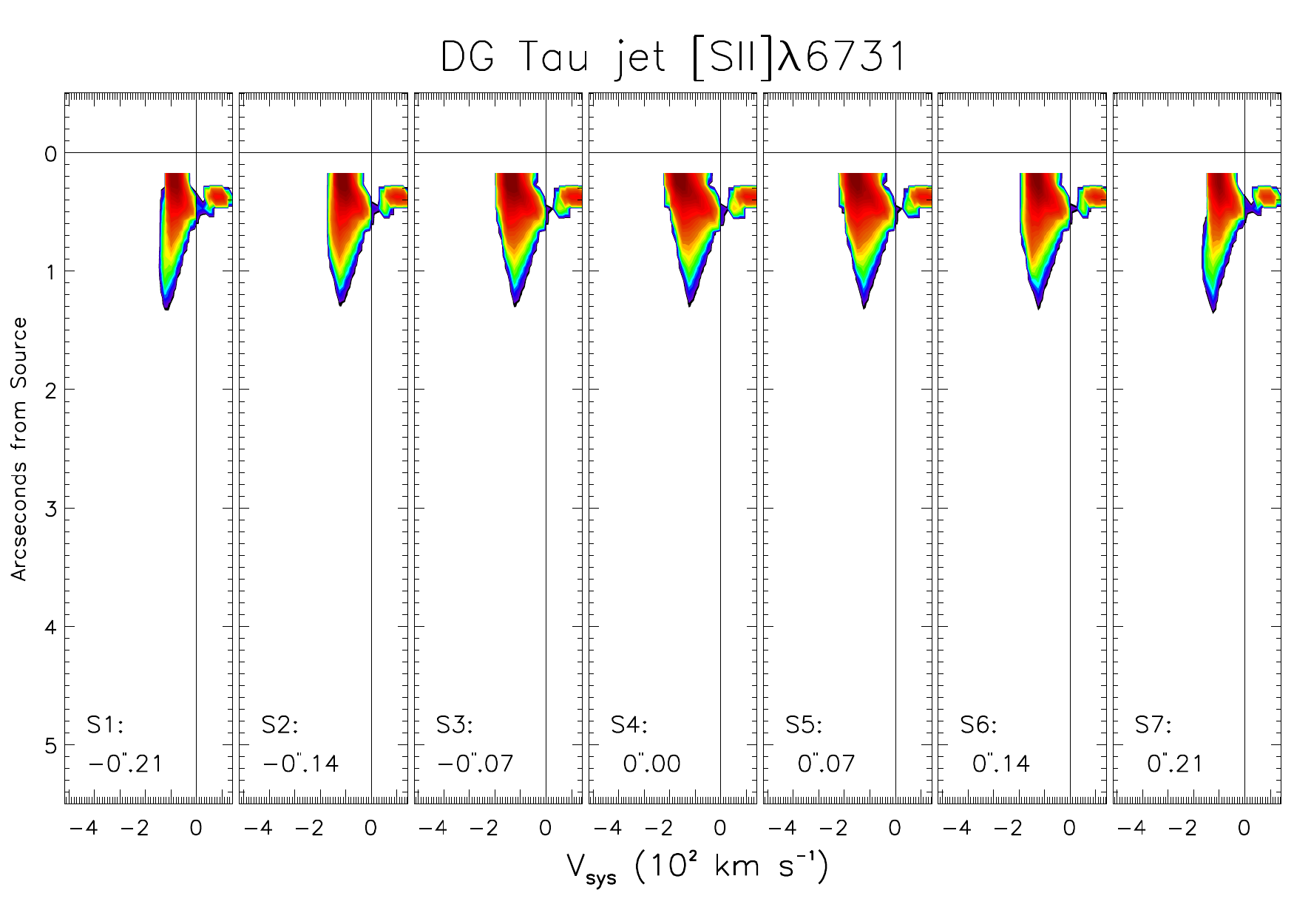}}
\caption{Synthetic PVDs obtained for the $\mu=-0.5$ case, with linear
tails added to the profiles in P06. Contour levels and reference lines are as in Fig.\,\ref{DG_Tau_SII}.}
\label{pvd_05_lin} 
\end{figure}

In wide-angle jets, a crucial role is played by tails. In the case of 
DG3 (Fig.\,\ref{emissivity_05_lin}), wider emitting cones 
are created by faster, linear tail-driven winds, which are able 
to push the hot gas much farther than the exponential tail-driven 
winds of the case of DG2 (Fig.\,\ref{emissivity_05_exp}). 
In both cases, the bifurcation 
of the red-emitting cone indicates the flow opening. 
The case of DG2 still shows an almost empty synthetic PVD, with a faint 
surface brightness, close to the nozzle, whose pattern is quite similar 
to that shown in Fig.\,\ref{pvd_0_exptail}.      
A richer kinematic structure appears in DG3, Fig.\,\ref{pvd_05_lin},
where the surface brightness and the velocity spread are comparable with the 
observed velocity spread in all slit positions. 

The emitting blob extends 
up to $\sim 1.5''$ from the nozzle, and in the central slit position 
radial velocities (with respect to the observer) of order 
200 km/s are observed. Lateral slit positions show lower 
velocities that increase when moving away from the source, 
in agreement with the features of the stationary 
first blob of DG-Tau (see Sect.\,\ref{observations}). 
Nevertheless, it is apparent that other mechanisms should be included
to reproduce all the properties of DG-Tau microjet. Namely, 
the second moving, emitting blob at $3-4''$, shown in
Fig.\,\ref{DG_Tau_SII} is probably due to 
temporal variations of the ejection properties. In order to reproduce it, in the next case 
a time-dependent component has been added to the steady component of 
the flow.
 
Finally, in observational PVDs the first blob shows a velocity gradient 
that might be due to either local (i.e., in time) particle deceleration 
at the nozzle position, or spatial particle acceleration across the blob region. 
This latter kind of gradient cannot be reproduced   
in our simulations, since spatial acceleration mechanisms are not at work 
in our model. On the other hand, local temporal fluctuations at the nozzle
can be reproduced by setting unsteady inflow conditions, as in the case of DG4.

\subsection{Simulations of the DG-Tau case: the time-dependent model} 
To reproduce the second moving blob in images of the DG-Tau jet, 
in this experiment (DG4) we added a time-dependent component
to the stationary flow used for DG3 ($\mu=-0.50$), 
regardless of the physical mechanism responsible for such a component.
Including temporal fluctuations introduces many new parameters:  
perturbation shape and strength, period (defined as the single event 
life-time) and periodicity (gap between two episodes).
Exploring the full range of these parameters is far beyond the scope 
of this experiment, whose sole aim is to add a variation in the 
ejection process capable to justify the moving, fainter knots at $3-4''$. 
At this scope, we adopted a saw-tooth 
velocity profile, where
the velocity perturbation linearly grows from 0 to 
$\sim v(r,z_\mathrm{nozzle})\times 0.9$ in one period of time, and
then abruptly drops to 0. Period and periodicity are set equal 
to 4 yr and 8 yr.

\begin{figure}[!htb]
\resizebox{\hsize}{!}{\includegraphics{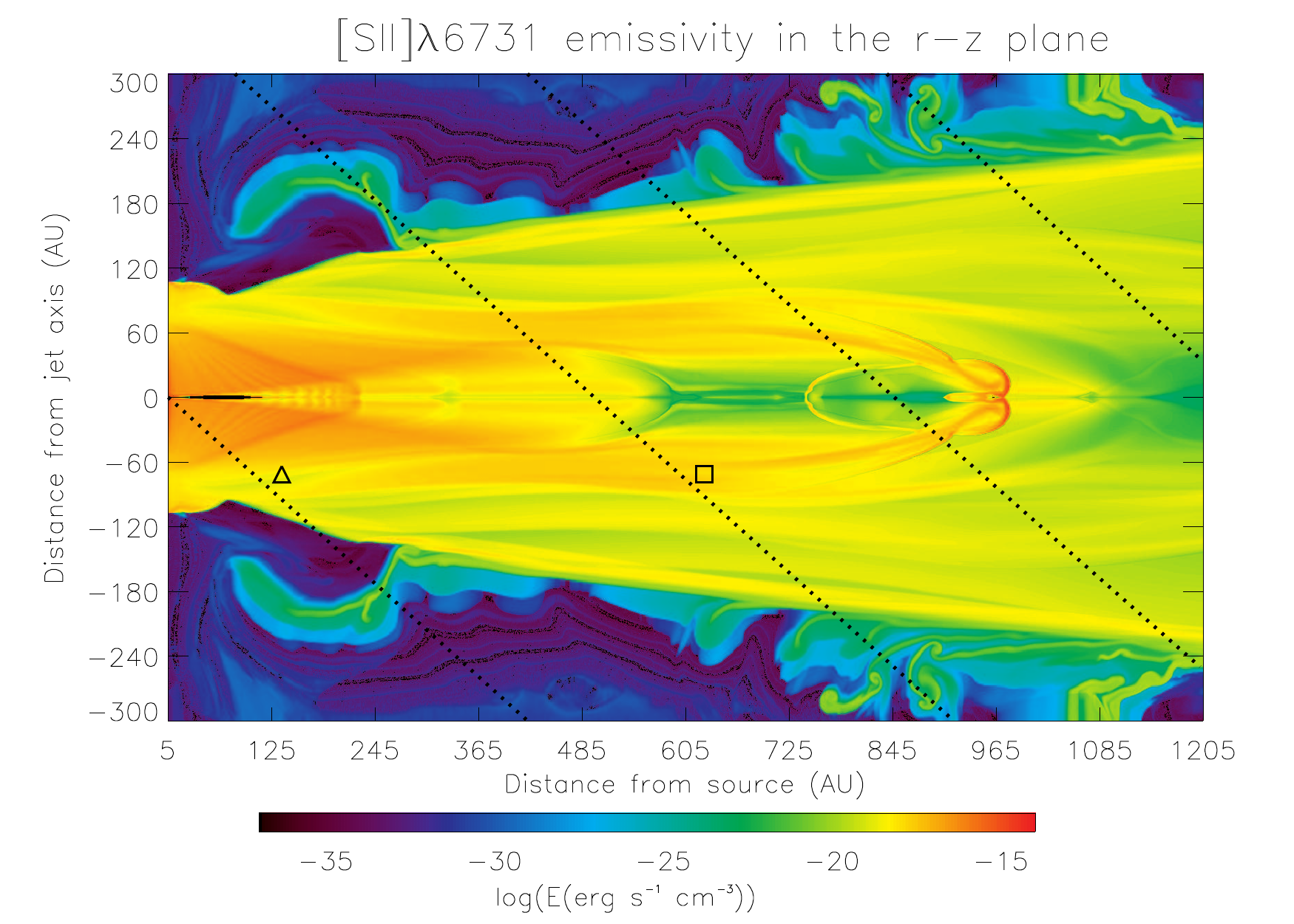}}
\caption{Emissivity map for the unsteady 
DG4 ($\mu=-0.5$) case. The dotted line marked with a triangle indicates 
the integration path along the line of sight, whose angle with the jet 
axis is 38$^{\circ}$. The lines marked with a square, at 
$\sim 4$~arcsec from the source, shows the position of the head of 
the second blob. } 
\label{DG_blob_2d}
\end{figure}

\begin{figure}[!htb]
\resizebox{\hsize}{!}{\includegraphics{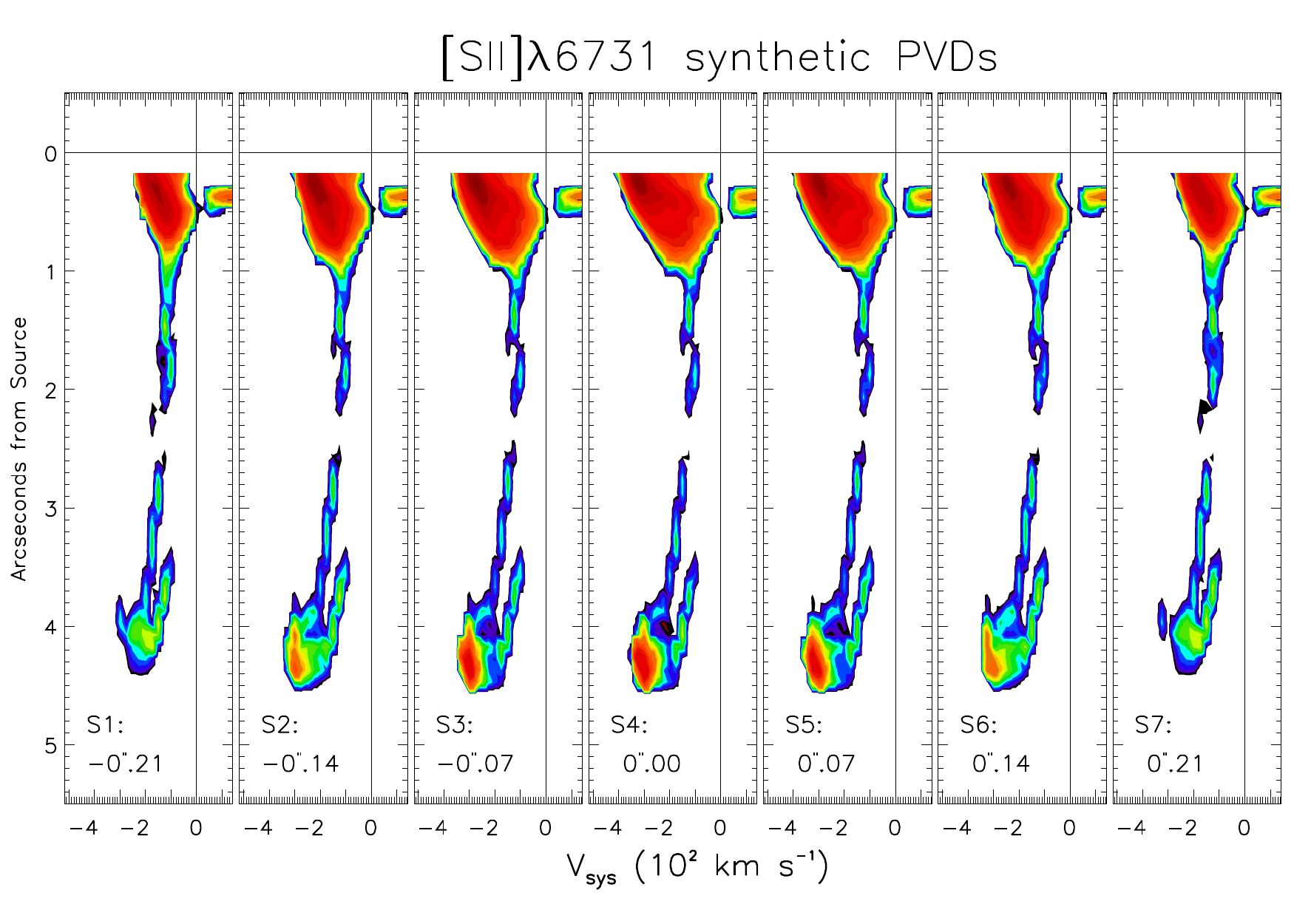}}
\caption{PVDs for the 
nonsteady $\mu=-0.5$ case (see also MA14).} 
\label{s2_non-steady}
\end{figure}

\begin{figure}[!htb]
 \resizebox{\hsize}{!}{\includegraphics{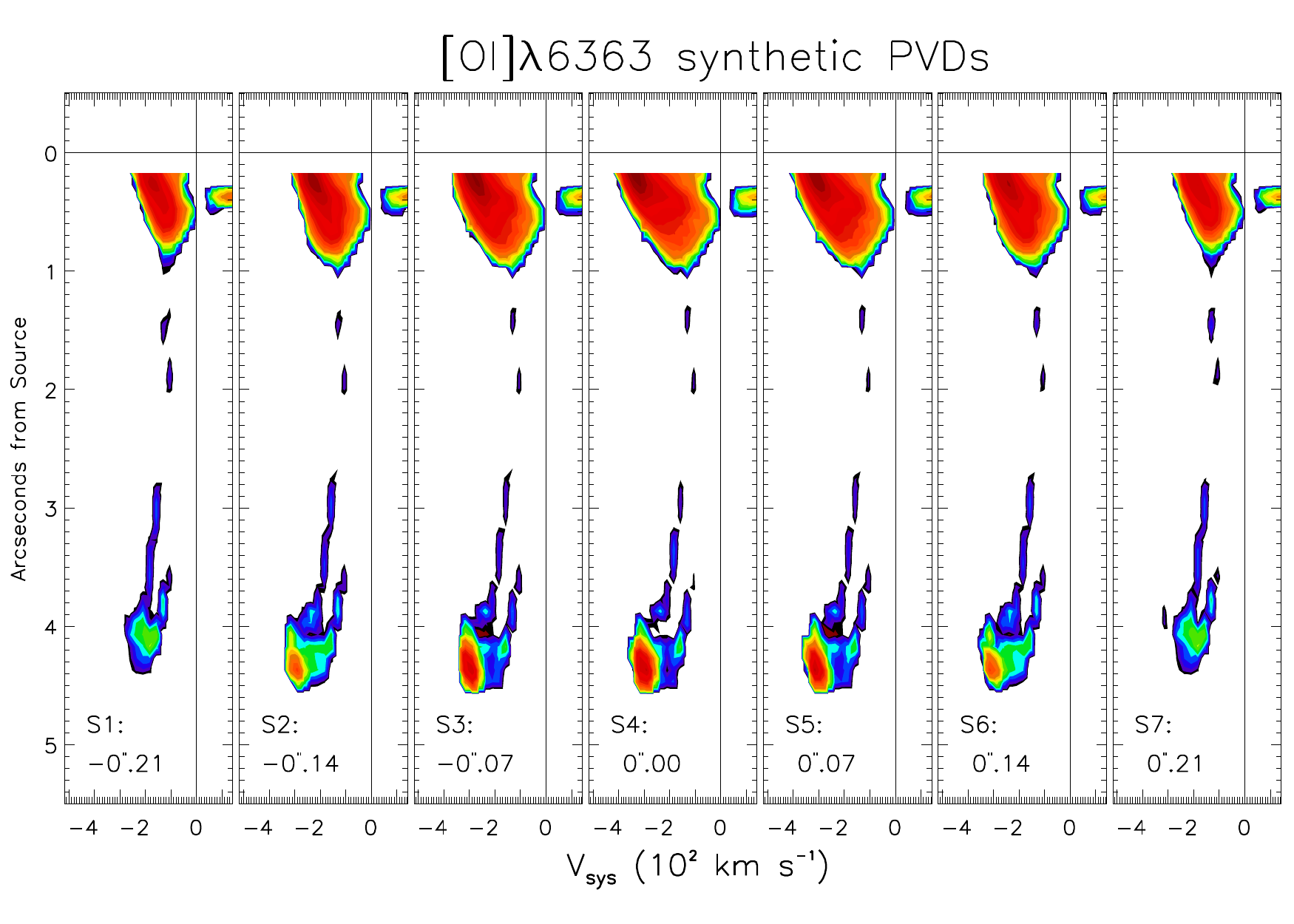}}
       \caption{Same as Fig.\,\ref{s2_non-steady} for a
different emission line.}
         \label{oi_non-steady}
   \end{figure}


Fig.\,\ref{DG_blob_2d} shows the 2-D emissivity map in 
[\ion{S}{ii}]$\lambda$6731, whereas synthetic PVDs
in sulphur and oxygen are shown in 
Figs.\,\ref{s2_non-steady}, \,\ref{oi_non-steady},
These figures reveal similarities and discrepancies with observations.
The blobs position is correct, 
in the range between $3-4''$ from the source, and the oxygen emissivity 
is weaker than sulphur emissivity as expected (see MA14). On the other 
hand, the spatial extension and the velocity spread are still too small. 
In particular, the second blob brightness is too strong, and positions from S2 to S6
show a red-emitting zone (in the figure colors palette) that is not present 
in observational images.

Discrepancies probably depend on the choice of the perturbation 
parameters. In particular, the saw-teeth fluctuation profile 
tends to steepen when propagating with the flow, the same way the 
solution of the nonlinear Burgers equation does, forming strong 
shocks near the nozzle. 
We are confident that a more exhaustive exploration of unsteady 
inflow conditions will lead to more realistic emitting patterns. 
The main result, however, is the robust link between synthetic PVDs and 
$\mu$ values. In fact, when the temporal fluctuations 
used in DG4 case are applied to models $\mu=0.0$ and $-0.25$ (case DG1), or to 
the $\mu=-0.5$ model with exponential tail of case DG2, quite different 
solutions are obtained, which are not at all comparable with observations.

\subsection{The RW-Aur jet} 
\label{rw_aur}

Cases RW1, RW2, and RW3 try to reproduce the features of the 
microjet from RW-Aur. At this scope, we used the values of the 
physical parameters of RW-Aur cited in Tab.\,\ref{inflowparam},
in particular, $n_\mathrm{H}(r_i)=2.4 \times 10^7$ 
cm$^{-3}$ and $B(r_i)=0.04$~G, which are 
different with respect to DG-Tau values. 
We run models for $\mu = -0.5, -0.25$ and $0.0$, respectively, by using 
linear tails to extrapolate the inflow radial profiles. 
In order to reproduce the moving chain of observed emission knots, 
time-varying inflow conditions are used, defined by 
a positive sinusoidal temporal profile
$$
V_\mathrm{pert} = A \sin^2 (2 \pi t/T),
$$
where the amplitude $A$ is 255, 190, 155 km~s$^{-1}$ 
for $\mu = -0.50$, $-0.25$ and $0.0$, respectively,
and $T=5$~yr, which yields the perturbation growth timescale 
to be $T_g = 2.5$~ yr,  a value that is close to the high-frequency 
perturbation period cited by \citet{Lopez03}. Amplitudes, in turn, 
have been assumed to be equal to the velocity peak values computed from 
the P06 models, see Fig.\,\ref{inflow_profiles}.

\begin{figure}[!htb]
 \resizebox{\hsize}{!}{\includegraphics[angle=180,origin=c]{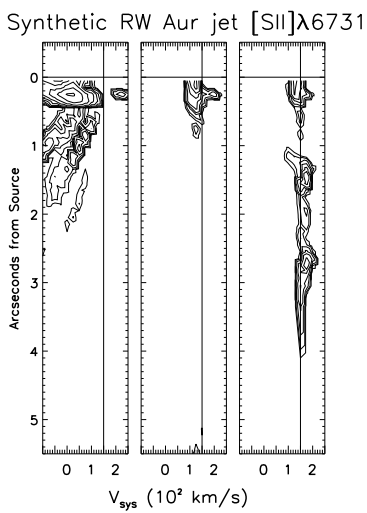}}
       \caption{Synthetic PVDs for S4, 
central slit position, and for three different models, 
$\mu=-0.5,\, -0.25,\, 0.0$ (from left to right).}
         \label{RW_3mu}
   \end{figure}

Results for the three models have been collected in Fig.\,\ref{RW_3mu}, 
where synthetic PVDs for the central slit position S4 are shown.
The figure reveals that models $\mu=-0.5$ and 
$-0.25$ are far from the morphological features of the microjet from
RW-Aur (\citet{Melnikov09}, Fig. 1, 
whereas $\mu=0.0$ provides better results. In particular, 
a chain of emitting knots appears in the first 5 arcsec, 
though the velocity spread is still smaller than the observed spread.  

These results suggest that the jet from RW-Aur might originate 
from outflows with collimation degrees even higher than that of
$\mu$ = 0. In this frame, we run the case of RW4, listed in 
Tab.\,\ref{runlist} as a super collimated (SC) outflow.

\begin{figure}[!htb]
 \resizebox{\hsize}{!}{\includegraphics{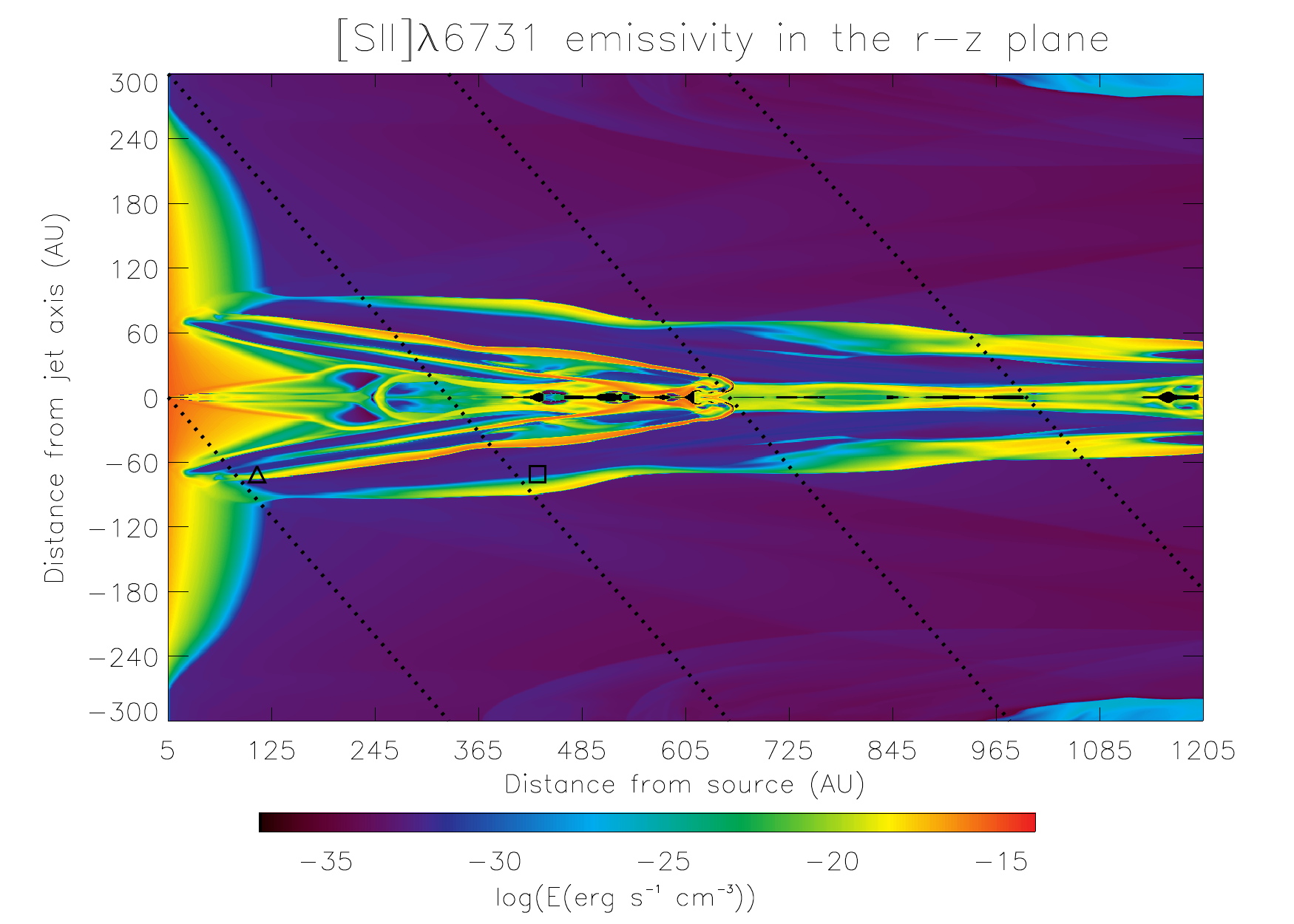}}
       \caption{Emissivity map for case RW3
     ($\mu=0.0$). Dotted lines show the integration path along the line of sight,
     whose angle with the jet axis is 46$^{\circ}$. The lines marked with 
     a triangle and a square correspond to positions $z=0$ and $\sim 3.4''$ 
     in the PVD.}
         \label{RW_2d}
   \end{figure}

We highlight that the inflow profiles of case RW4 do not 
correspond to any model from P06. Instead, by following the gross behavior
of the profiles of Fig.\,\ref{inflow_profiles} when moving from 
larger to smaller (negative) values of $\mu$, or from less collimated to 
more collimated outflows, we calculate the profiles
by doubling the intensity of the magnetic field and halving  
the longitudinal velocity with respect to the $\mu$=0 case. 

\begin{figure}[!htb]
 \resizebox{\hsize}{!}{\includegraphics{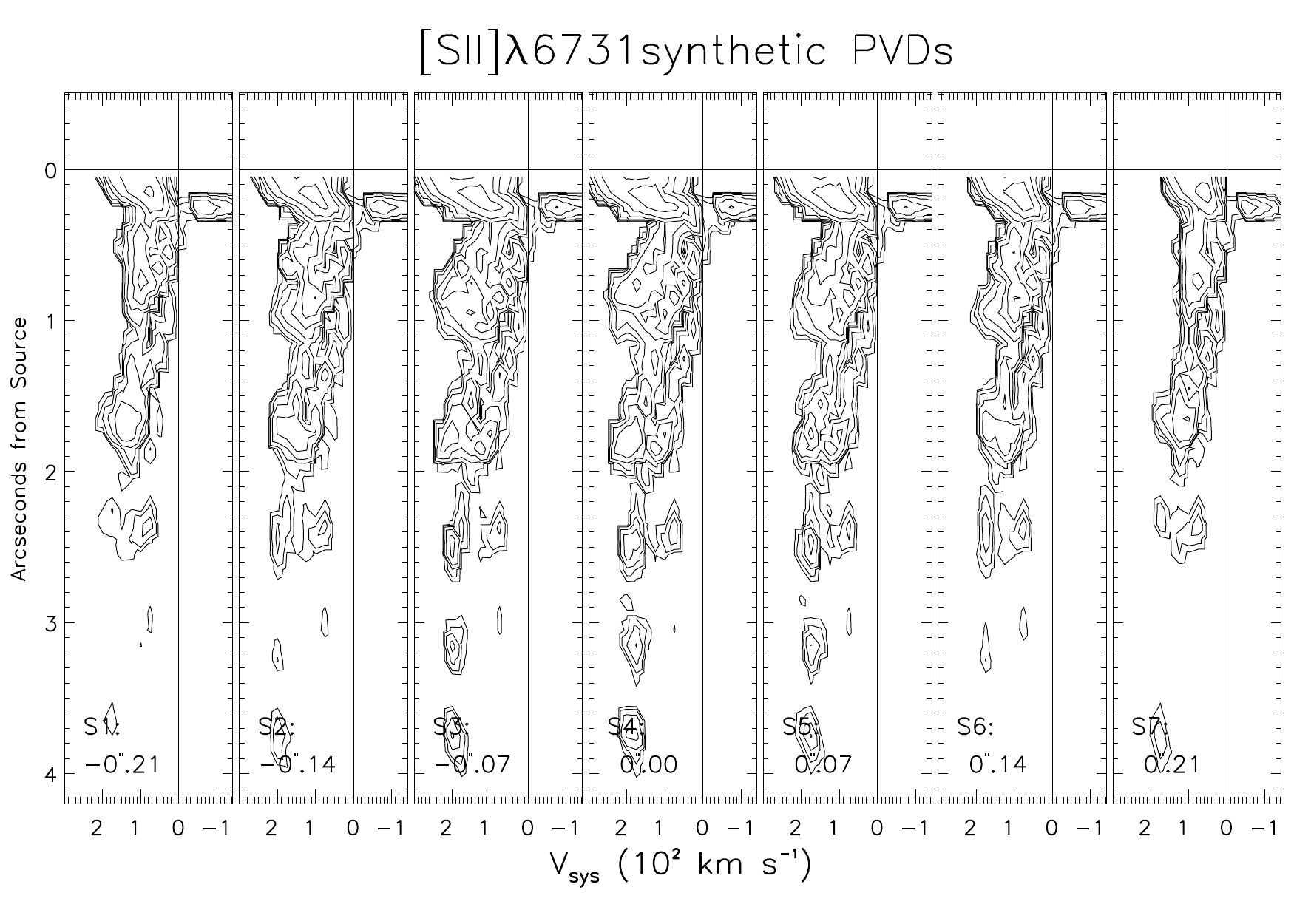}}
 \caption{Results for the RW4 case, corresponding
 to the SC run, see for comparison 
\citet{Melnikov09}, Fig. 1.
 }
         \label{RW4}
   \end{figure}

Results for this case are shown in Fig.\,\ref{RW_2d} and Fig.\,\ref{RW4}.
Fig.\,\ref{RW4} can be directly compared with the observed PVDs of 
Fig.~1 \citep{Melnikov09}, since velocity interval and scale are 
the same.

Synthetic PVDs reveal a chain of emitting knots, 
which are also visible in the 2D emissivity map of Fig.\,\ref{RW_2d}, 
whose morphology and kinematical features are in good agreement 
with the observed PVDs .
Results of pulsating simulations, for both 
DG4 and RW4 cases, strictly rely on the perturbation shape and 
growth timescale, 
(4 and 2.5~yr, respectively). Changing these parameters yields
quite different solutions. 
We stress that investigating the pulsating inflow conditions 
parameters is beyond the scope of this paper. Temporal fluctuations 
parameters of cases DG4 and RW4 were chosen according to the following criteria:

\begin{enumerate}

\item We chose the fluctuation growth timescale, $T_g$, in the range 
of the values reported in Sect.~\ref{observations}. In particular,
we adopted a small timescale perturbation, with $T_g=2.5$~yr for the case of RW4
\citep{Lopez03}, and $T_g=4$~yr for the case of DG4 \citep{Pyo03}.
\item For both DG4 and RW4, we avoided sharp discontinuities to suppress strong post bow-shock emission lines, in favor of smooth 
saw-tooth or sinusoidal temporal profiles. 

\end{enumerate}

The aforementioned choices have proved to achieve better results in terms of 
comparison between synthetic and observed PVDs.

\section{Discussion and conclusions} 
\label{conclusions}

We performed numerical simulations of axisymmetric 
magneto-hydrodynamic jets with the PLUTO code. First, a supersonic outflow, 
injected from a nozzle placed on the left boundary, sweeps 
the numerical domain, until the heading bow-shock leaves
the right boundary. Then, the simulation continues with a jet flowing in the domain. 
Initial and boundary conditions were retrieved 
by using a mix of observational constraints and theoretical models.

In particular, the nondimensional radial profiles of the main 
physical quantities at the nozzle position are taken 
from \citet{Pudritz06}. In this paper,  
self-collimating outflows have been generated via numerical simulations 
at the distance of $\sim 5$~AU from the disk. 
Each solution corresponds to a different value of the 
parameter $\mu$ in the relation that rules 
the behavior of the poloidal magnetic field on the disk surface:

\begin{equation}
B_z(r_0,0)\propto r_0^{\mu-1}
\label{initialfield2}
\end{equation}

In particular, we consider four different values of this parameter: 
$\mu=0.0, \,-0.25, \,-0.5, \,-0.75$. 
The nondimensional profiles are converted into dimensional profiles, 
with proper scale factors. Since the inflow profiles 
from P06 extend for a few AUs from the jet axis, linear or exponential 
extrapolations (the tails) are used to match the inflow 
conditions and the ISM.
Finally, the inflow temperature and ionization fraction profiles are 
chosen according to observational criteria.
The numerical solutions obtained by the PLUTO code are finally
used by the Optical Telescope Simulator post-processing code,  
and synthetic PVDs of optical forbidden emission lines are obtained.
The comparison between synthetic and observational PVDs 
allows us to check the reliability of the inflow conditions at the 
nozzle and, in turn, provides feedback on the magneto-centrifugal 
launch models used to generate the conditions themselves.
In this preliminary work, simulations and synthetic PVDs extend over 
the first $5''$, since we focus on the microjets from DG-Tau and RW-Aur,
that have bright emission over that distance from the star.

Our main results are listed below:

\begin{itemize}

\item[-] The setup procedure shown in Sect.~\ref{setup} enables us
to match numerical simulations in the launching and propagation region.
Under the prescriptions given in the Acceleration-Propagation 
Matching (APM) procedure, magnetized fluids that are numerical solutions 
of magneto-centrifugal models inside the accelerating region,
are allowed to flow into the (observable) propagation region. 
Synthetic PVDs of optical forbidden lines 
are returned by the OTS code, and the surface brightness in various velocity channels 
are reconstructed with the same resolution used by the HST/STIS spectrograph.

\item[-]  Results from P06 show that different 
radial profiles of the magnetic field at the disk surface, 
characterized by different values of the parameter $\mu$,
lead to accelerated outflows with different 
degrees of collimation. In particular, cases $\mu=0.0,\, -0.25$ yield 
collimated outflows, while $\mu= -0.50, \,-0.75$ lead to wide-angle jets, 
with a hollow cylinder in the central region. 
We show that such behavior is maintained in the propagation region.

\item[-] The synthetic PVDs of the jet from DG-Tau show that 
inflow conditions corresponding to collimated outflows 
in the acceleration region ($\mu=0.0,\,-0.25$) are not consistent with 
observations. In fact, high-velocity components are not present in
lateral slit positions, and the surface brightness is 
dominated by the oblique shocks surrounding the jet axis.
The 2D emissivity maps of the case $\mu= -0.50$, however, reveal a lack of internal 
shocks and a weaker beam component. When linear tails are used to extrapolate 
the radial profiles at the nozzle, the obtained synthetic PVDs are 
in good agreement with the observed profiles. Linear tails, in fact,  
increase the surface brightness in lateral slit positions, and  
reproduce the observed low-velocity lateral winds, close to the star. 

\item[-] The jet from RW-Aur can be better reproduced at a quality level
by inflow conditions corresponding to collimated 
outflows. The profiles derived by the case $\mu$ = 0, in fact, 
produces synthetic PVDs that show the gross features
of the observed profiles: namely, the knotty structure and the narrow velocity 
spread, centered around the value 100 km~s$^{-1}$, as in Fig.~1
 \citep{Melnikov09}. Even better results are obtained with 
inflow profiles not available in P06 (i.e., our SC, Super Collimated, case).
The model $\mu = -0.50$ yields wide-angle jets in 
the propagating region, where the flow unfolds downward from the nozzle,
and the emissivity features do not match observations. 

\item[-] Finally, to reproduce the features in observed PVDs of 
both jets, it is necessary to add temporal perturbations 
to the steady solutions computed by the theoretical 
acceleration models. In particular, a saw-tooth time-dependent 
velocity component was added in DG-Tau simulations, whereas sinusoidal 
perturbations of the velocity field were used for RW-Aur. In both cases, 
the perturbation amplitude is of the order of the steady velocity field,
and the period is of a few years.
This kind of perturbation reproduces the gross features of the observed 
emitting blobs, though discrepancies are still present. 
Time-dependent simulations, however, yield a wide range 
of free numerical parameters to be set, and the exhaustive exploration of this parameter space
is far beyond the aim of this work. 

\end{itemize}

In conclusion, we have shown that the disk-wind model can lead to completely
different kinds of jets, suggesting that the mechanism could
be at the base of all observed stellar jets, possessing quite different properties. 
In particular, the jet from DG-Tau is better reproduced by 
inflow conditions derived from wide-angle disk-wind simulations, 
($\mu= -0.50$), surrounded by low-velocity winds, whereas the 
model for collimated outflows  ($\mu=0.0$) reproduces  
the features of the jet from RW-Aur much better.
In both cases, steady simulations are not able to 
reproduce all the observed features, and supplementary  
time-dependent components of the inflow conditions must be included. 

\begin{acknowledgements}
The authors wish to thank Claudio Zanni (Osservatorio Astronomico di Torino,
Italy), Silvano Massaglia, and Andrea Mignone (Universit\`{a} di Torino, 
Italy) for very useful and friendly discussions, and the
referee for helping us to improve the paper with pertinent comments.
We acknowledge the CINECA Award IsC04\_INJECTS, 2011 for the availability 
of high-performance computing resources and support.
\end{acknowledgements}

\bibliographystyle{aa}

\appendix
\section{The test case} 
\label{Test_case}

The simulations discussed in Sect.~\ref{results} were run 
in two steps. In the first, stationary phase,
the bow-shock sweeps out the ISM that fills the numerical box, 
up to 300 AU and 1200 AU in radial and longitudinal direction, respectively. 
Once it has left the numerical domain from the right side of the box, 
and the solution inside the domain has achieved a quasi-stationary 
configuration, the next pulsated phase is switched on,
and the emissivity properties of the inner region, close to the source,
are investigated.

In the present Appendix, the effects of the adopted numerical domain size and 
cooling model are discussed in some detail. For what concerns the
cooling model, in particular, the nonequilibrium SNeq cooling model was used, 
which takes hydrogen ionisation-recombination effects
and the corresponding cooling losses into account, under the prescription 
that temperature does not exceed 75000 K.

\begin{figure*}[th]
\centering
\subfigure{}\includegraphics[width=18.cm]{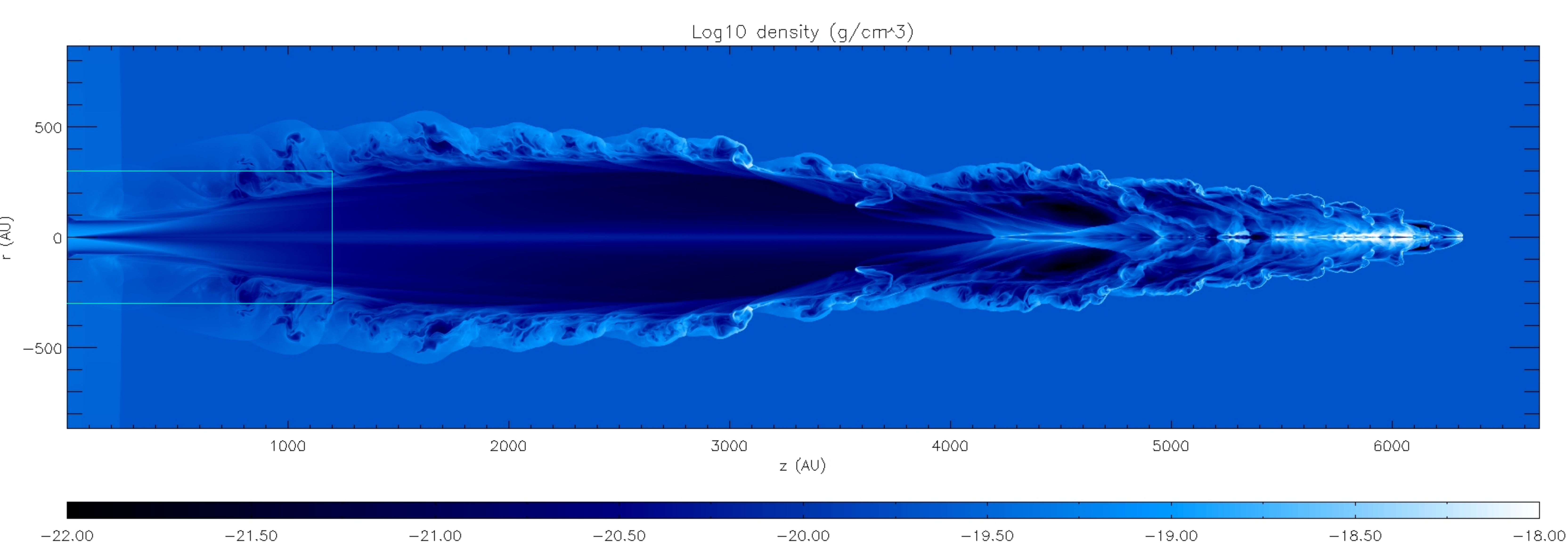}\quad%
\subfigure{}\includegraphics[width=18.cm]{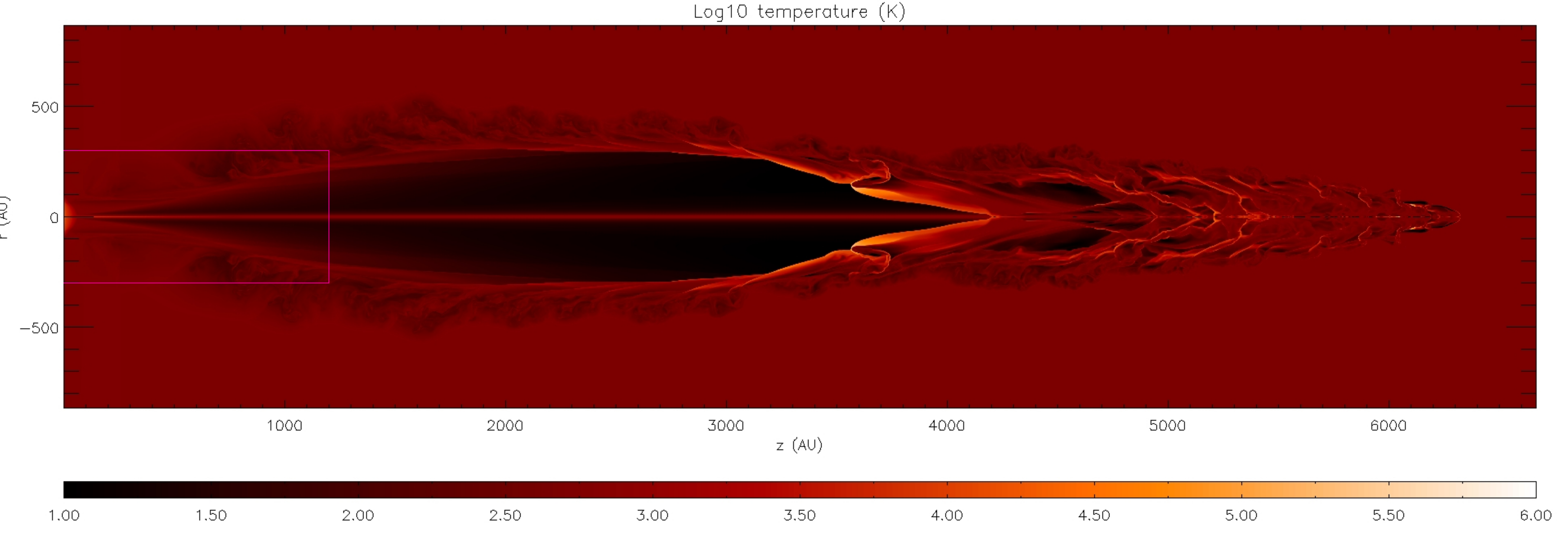}\quad%
\caption{Selected outputs from a large-scale simulation (Test$_{DG3}$). 
Solid lines in the nozzle region represent the area 
investigated in all simulations listed in Tab.\,\ref{runlist}.}
\label{Test_DG3_rho}

\end{figure*}

\begin{figure*}[tbp]
\centering
\hspace{-0.5cm}
\subfigure{}\includegraphics[width=7.5cm]{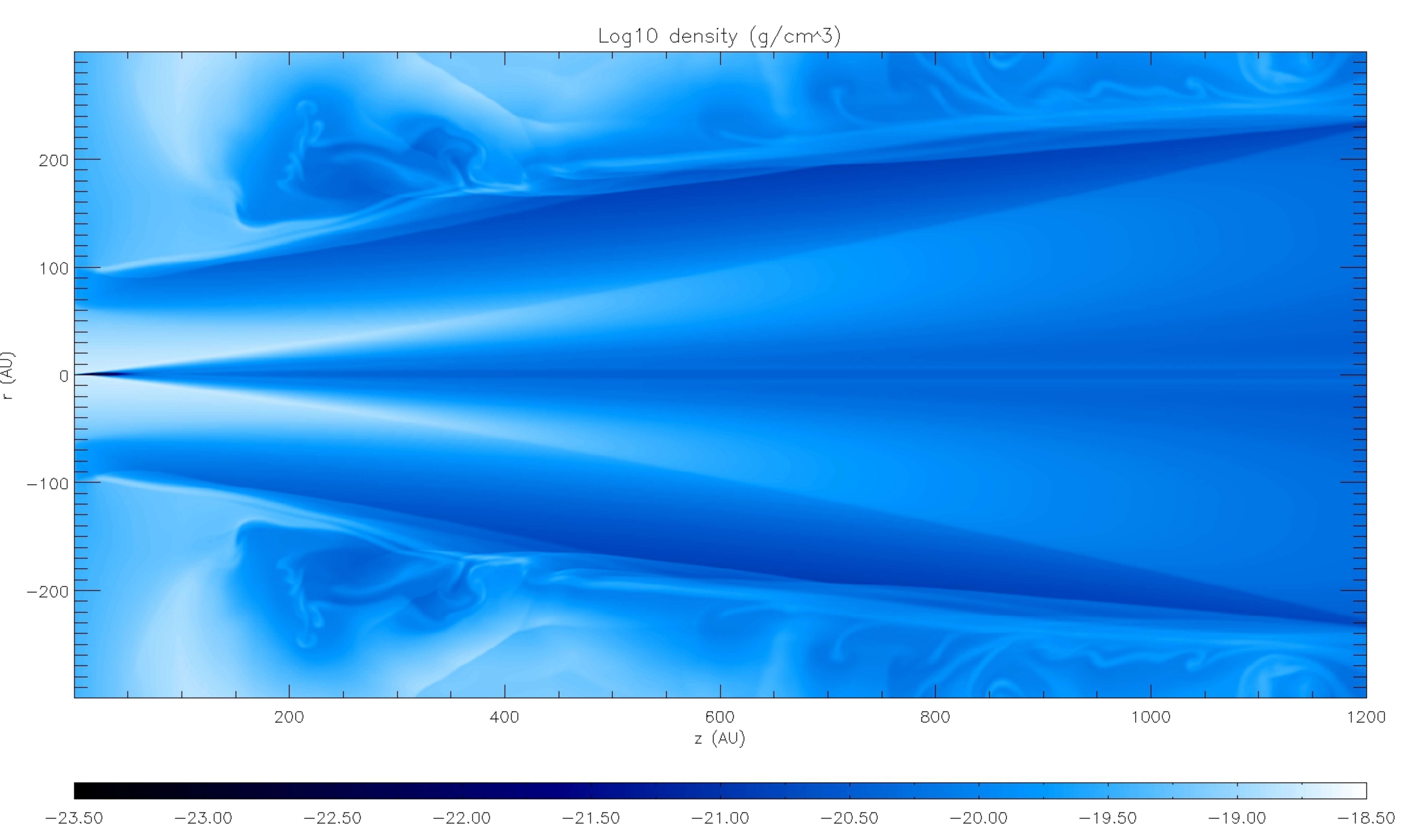}\quad%
\hspace{-0.5cm}
\subfigure{}\includegraphics[width=7.5cm]{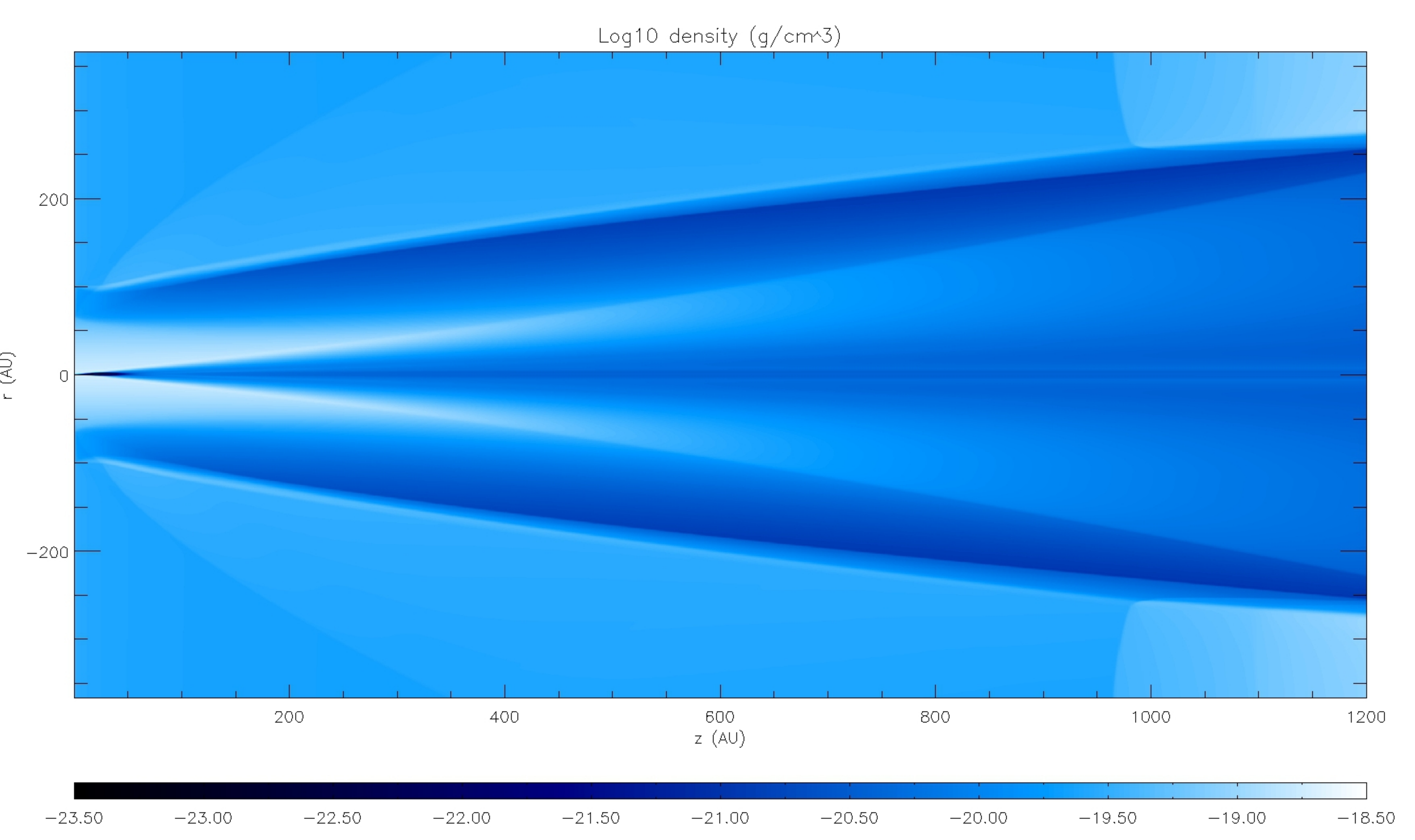}\\%
\hspace{-0.5cm}
\subfigure{}\includegraphics[width=7.5cm]{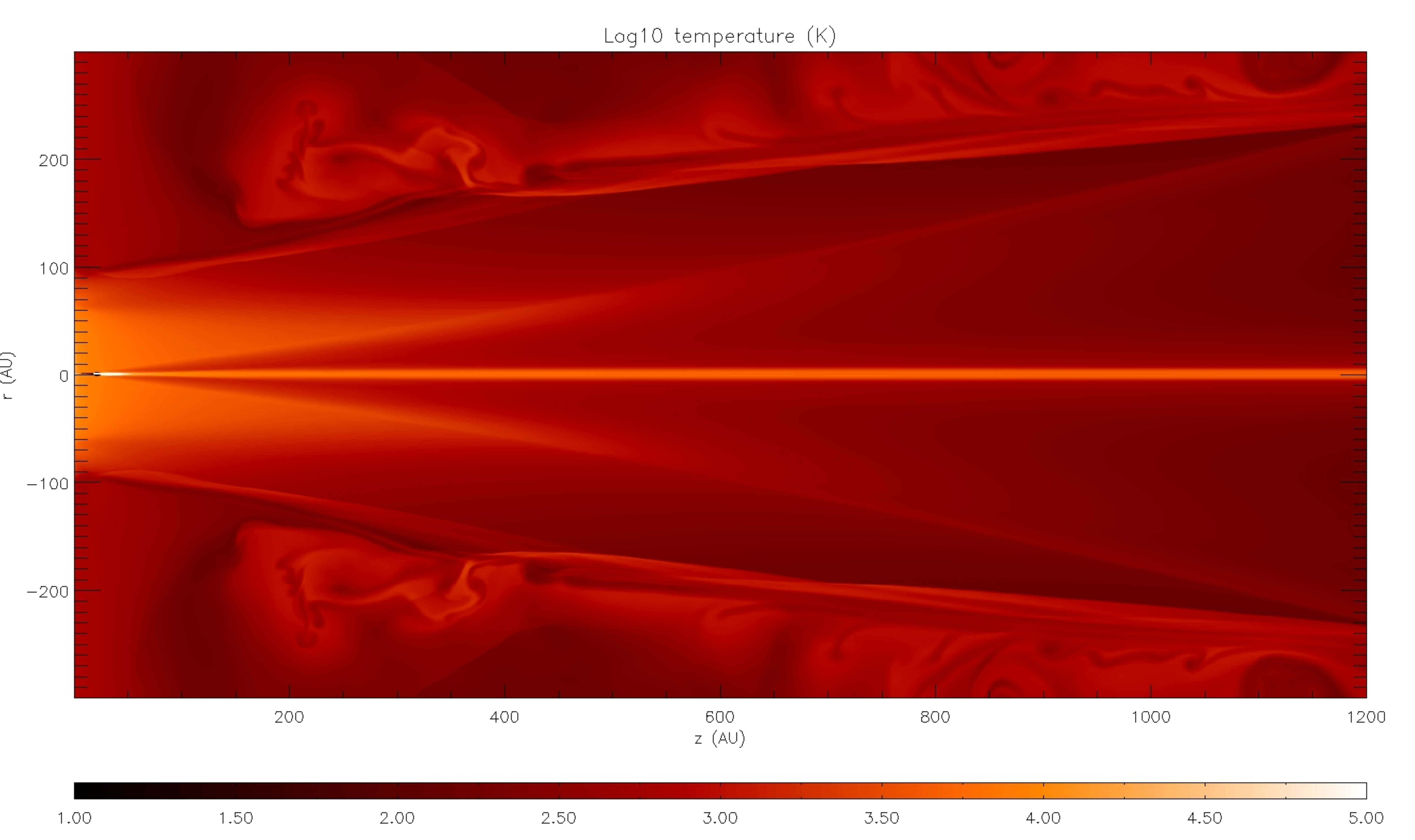}\quad
\hspace{-0.5cm}
\subfigure{}\includegraphics[width=7.5cm]{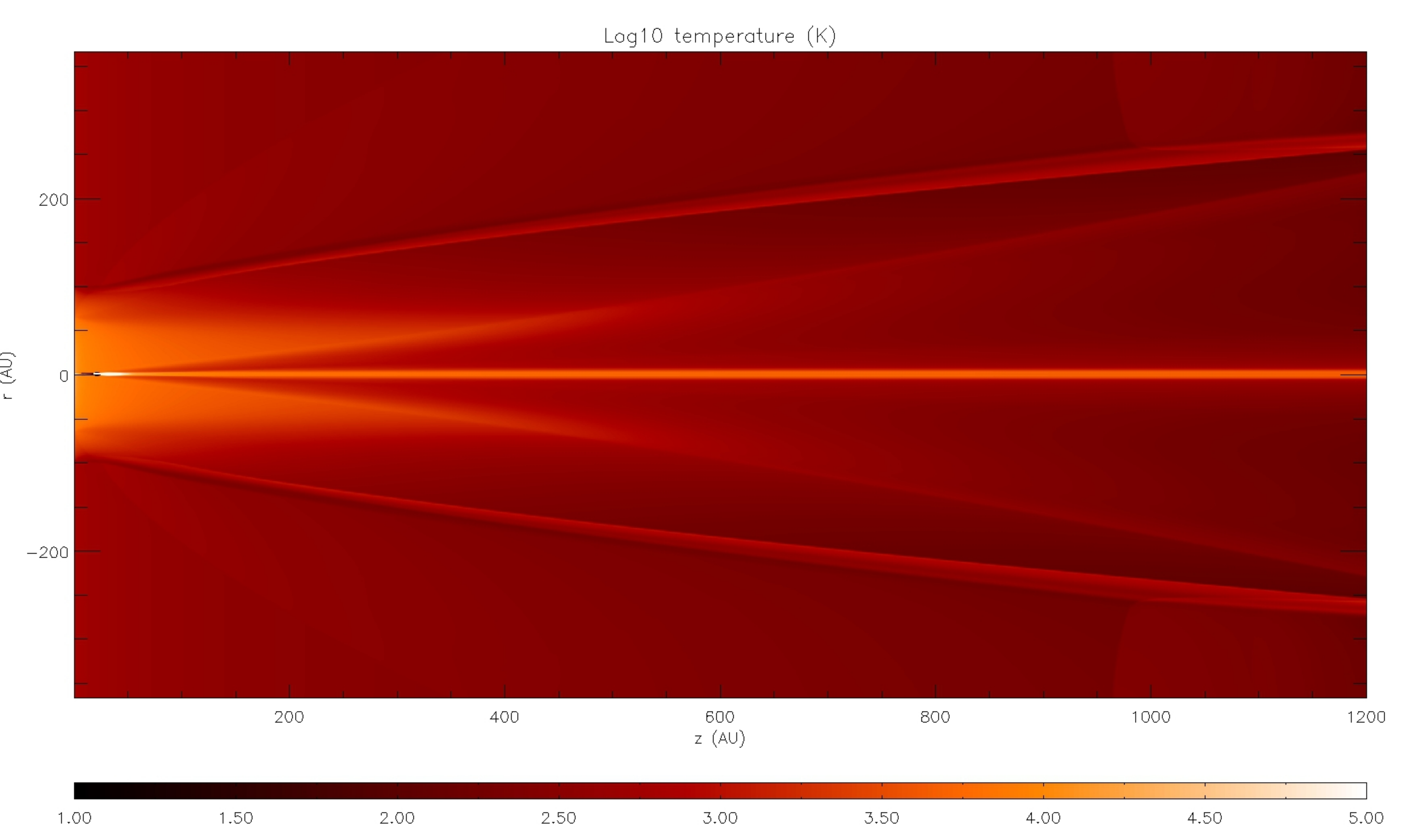}%
\caption{Density (top) and temperature (bottom) fields: 
the DG3 case patterns (left) {\it vs.} the stationary test case ones 
(right).}
\label{Test_rhosmallbox}
\end{figure*}

\begin{figure*}[tbp]
\centering
\hspace{-0.5cm}
\subfigure{}\includegraphics[width=7.5cm]{emissivity_05_lin}\quad%
\hspace{-1.3cm}
\subfigure{}\includegraphics[width=7.5cm]{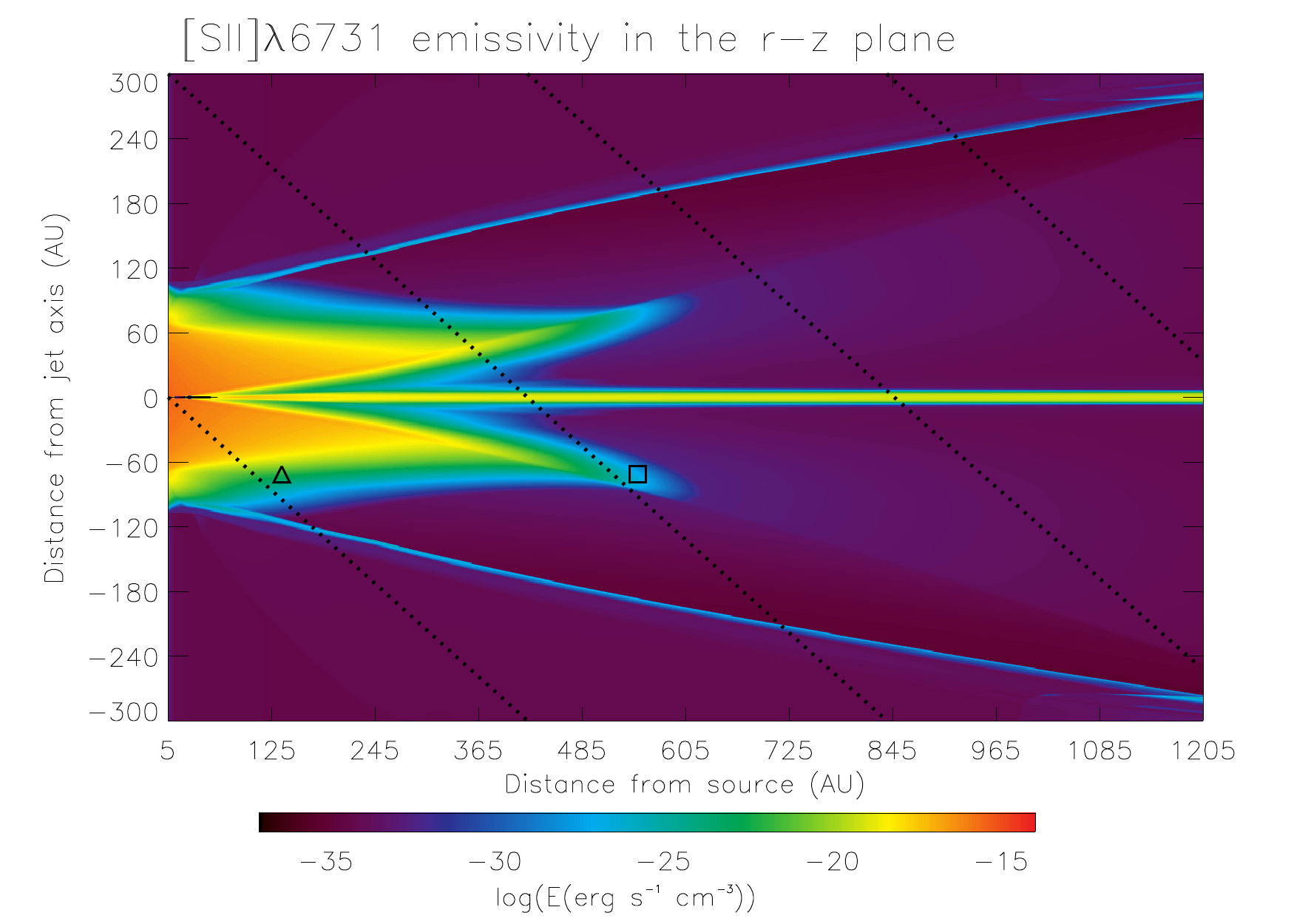}\\%
\hspace{-0.5cm}
\subfigure{}\includegraphics[width=7.5cm]{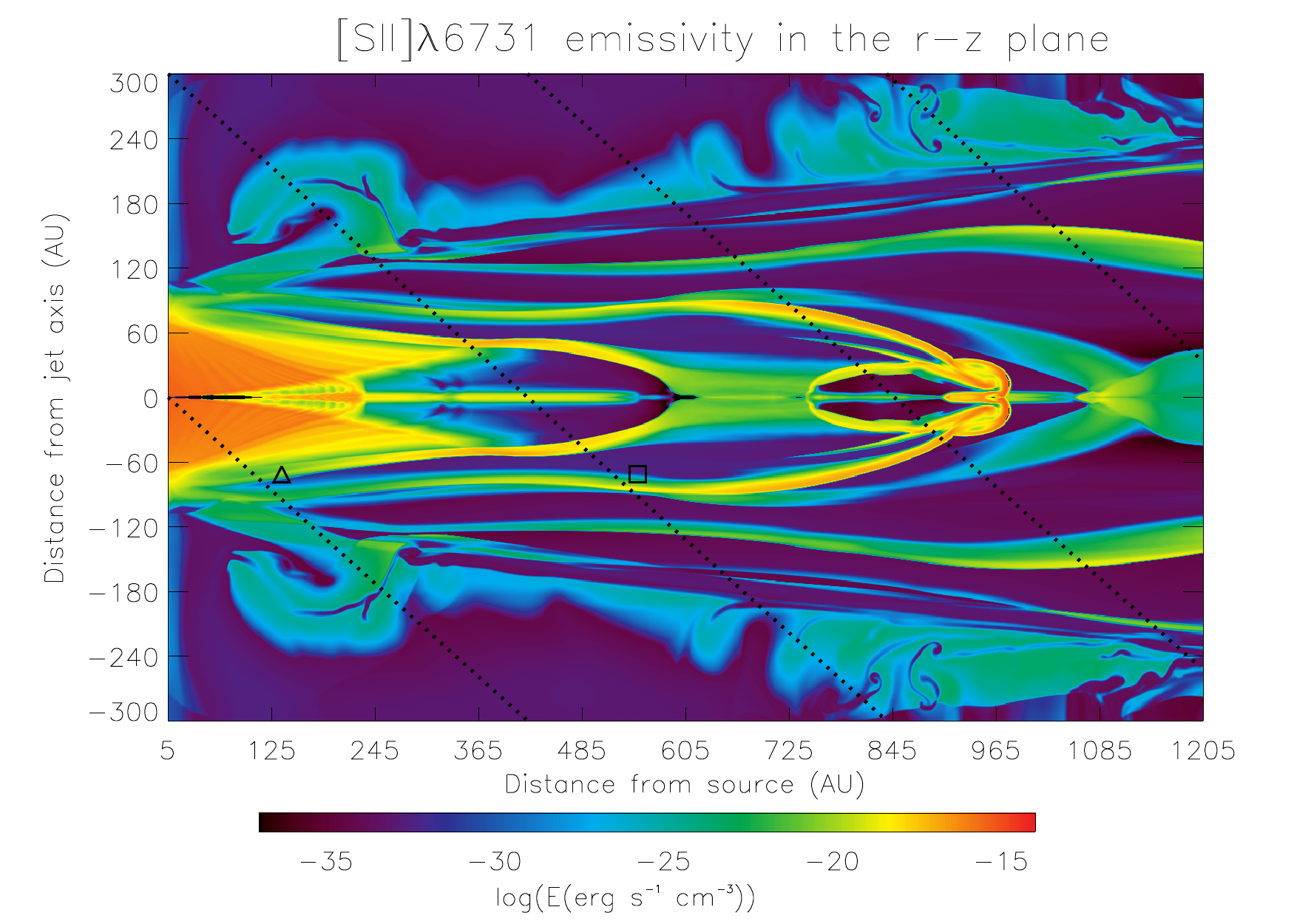}\quad
\hspace{-1.3cm}
\subfigure{}\includegraphics[width=7.5cm]{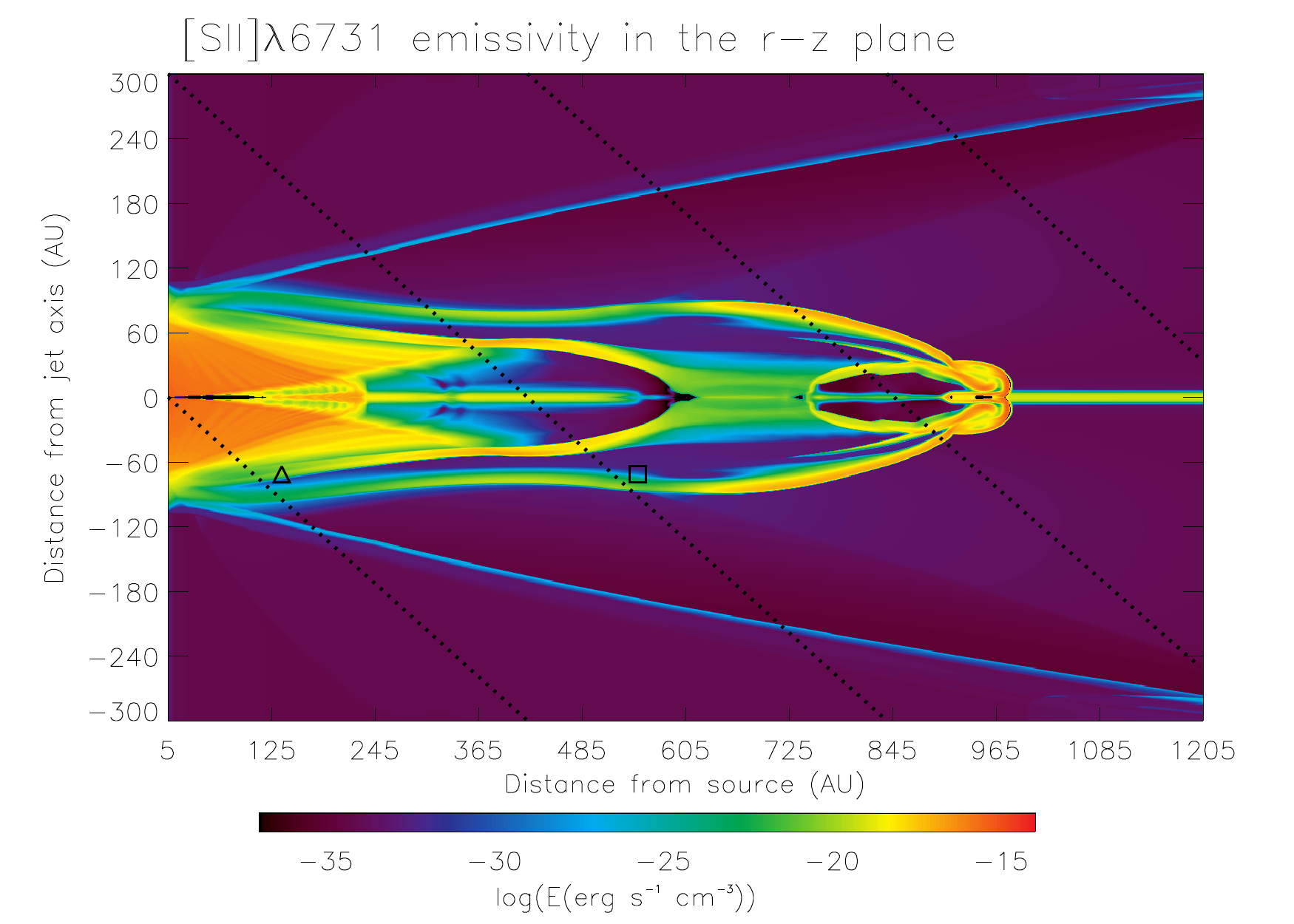}%
\caption{Two-dimensional emissivity patterns: the DG4 case (left) 
{\it vs.} the test case (right). The figures show both the stationary 
configuration (top) and the pulsated one (bottom).}
\label{Test_emiss}
\end{figure*}

\begin{figure*}[th]
\centering
\subfigure{}\includegraphics[width=8.5cm]{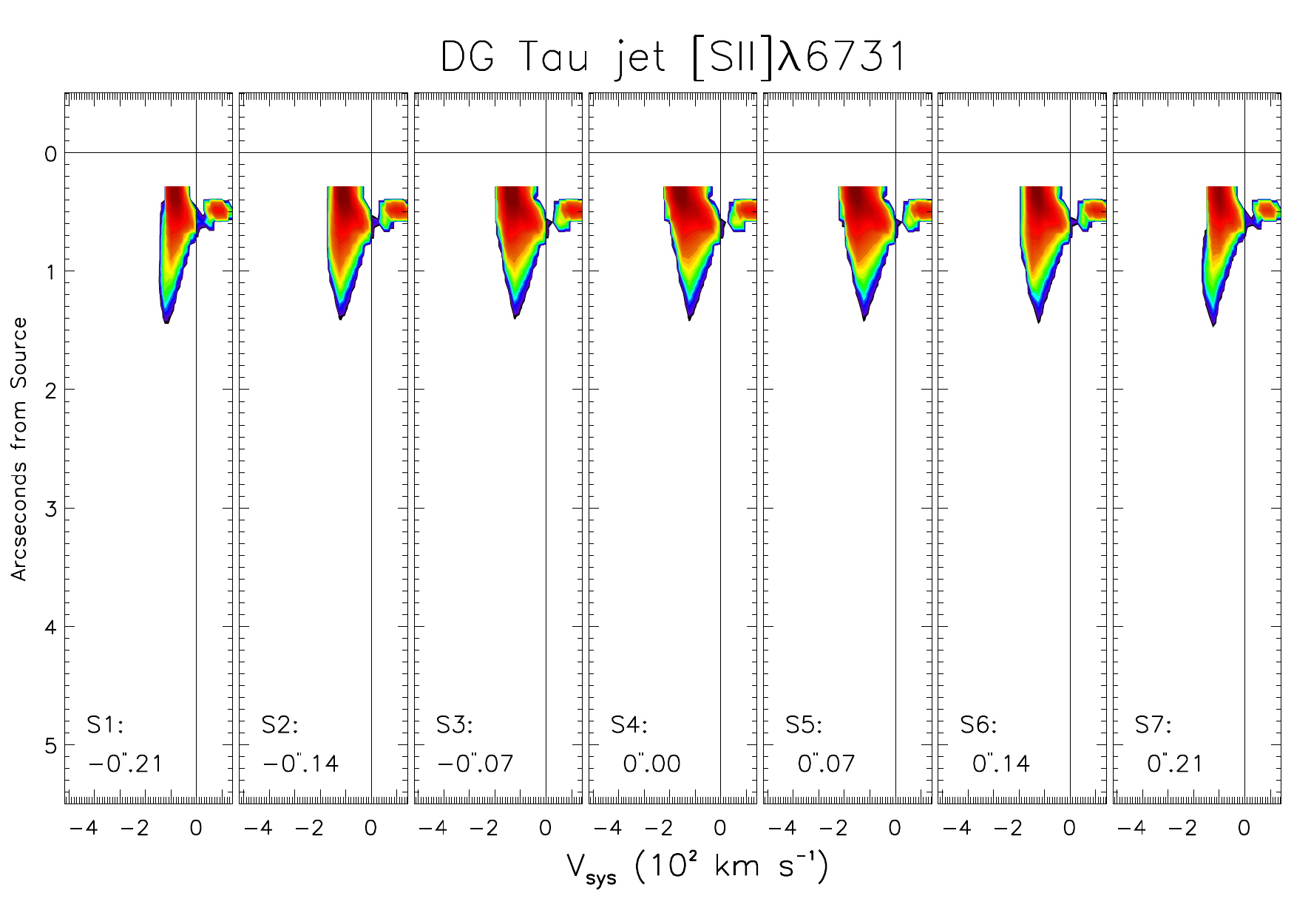}
\hspace{0.5cm}
\subfigure{}\includegraphics[width=8.5cm]{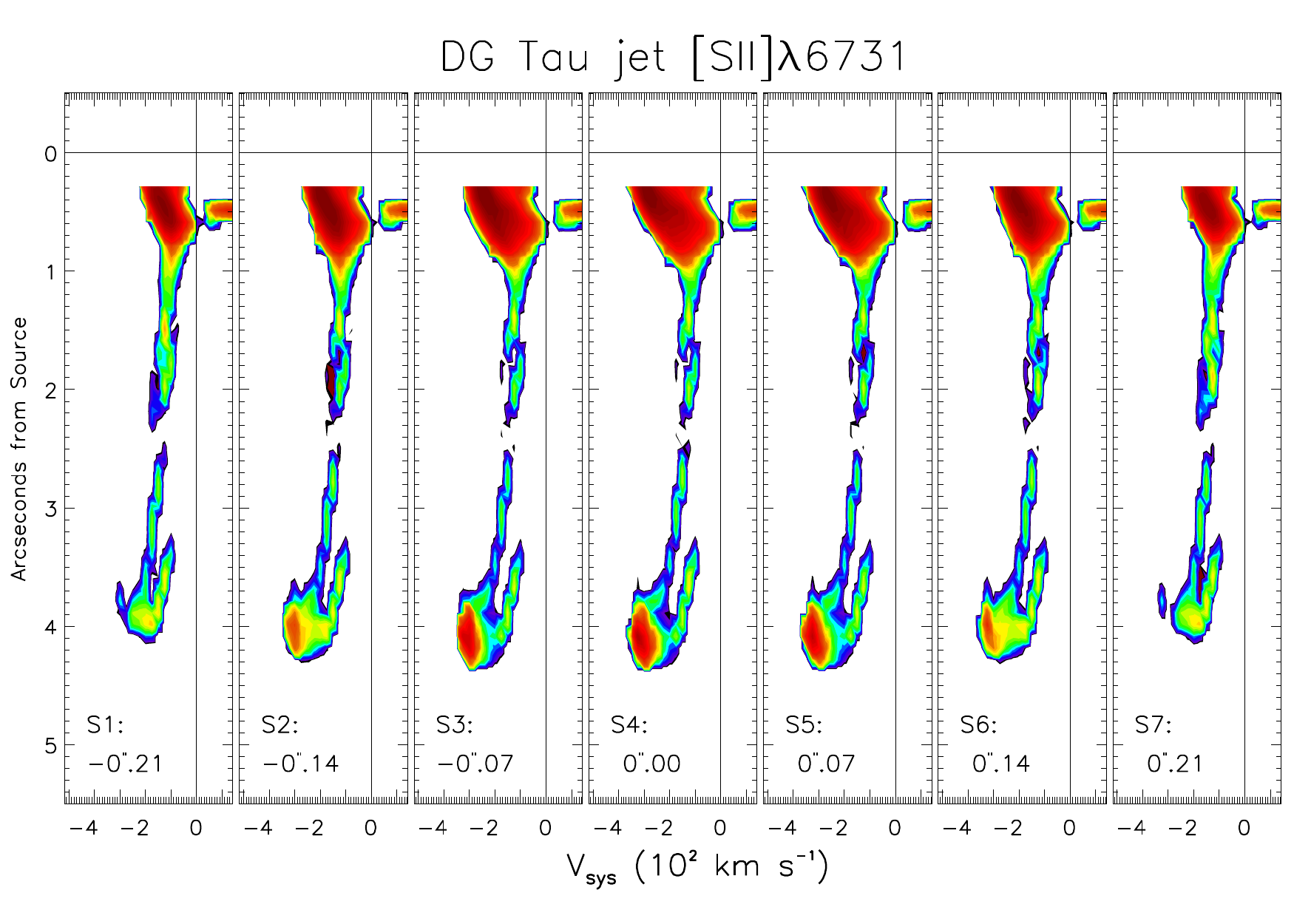}\\%
\caption{Left: synthetic PVDs for the
stationary $\mu=-0.5$ test case, that is simulation Test$_{DG3}$, to be compared
with Fig.~\ref{pvd_05_lin} of case DG3. Right: same for the unsteady 
simulation of case Test$_{DG4}$, to be compared with Fig.~\ref{s2_non-steady}. }
\label{Test_PVD}
\end{figure*}

Two questions arise: 
\begin{enumerate}
\item It is well known that bow-shocks (namely, the triple-point 
Mach-disk region behind them) are suitable to generate recirculating 
flows of matter that can move in reverse, and that they affect the region 
close to the nozzle. If so, how can we be sure that expelling 
the bow-shock from the simulation box at {\it only} 1300 AU from 
the source does not alter dynamical and emitting properties 
of the region under investigation, the {\it emitting region}, 
extending from the nozzle up to the first $\sim$5 arcsec?
\item  Though in the emitting region weak shocks are at work, 
with temperatures of order a few thousands degrees, well below
the cooling model limit, it is well known that in the post bow-shock region
temperature may achieve values of order 10$^6$ K.
If so, how can we be sure that miscalculating the cooling 
losses in the post-bow shock region does not affect the global 
jet dynamics and the properties of the emitting region as a consequence?
\end{enumerate}

In order to be confident with the results presented in the paper, 
we run a three-step test-case. In the first step,
a stationary jet propagates into a wider, large-scale domain
that extends over 870 x 6700 AU in $r$ and $z$ respectively, 
(1340 x 8060 grid points). Fig.~\ref{Test_DG3_rho} shows 
density and temperature fields (top and bottom, respectively) in the  
domain used in this large-scale simulation (to make a comparison, 
the small-scale domain of the runs of Tab.\,\ref{runlist} 
is highlighted in the figure with solid lines).

The adopted, simplified tabulated cooling model, in which  
cooling losses are a function of temperature and assigned (solar) 
abundances only, allows us to skip the temperature limit of the Sneq model. 
Thanks to both the larger domain size and the cooling model, the 
quasistationary solution obtained in this first step, once the 
bow-shock has left the domain, guarantees a better
description of the jet global dynamics, with respect to the simulations
of Tab.\,\ref{runlist}. 

In the second step (simulation Test$_{DG3}$), the internal portion 
of the stationary solution of the first step, marked in blue in 
Fig.~\ref{Test_DG3_rho}, is used to restart a stationary, 
small-scale simulation in which the precise Sneq cooling model 
is recovered to properly calculate the emissivity pattern 
and synthetic PVds.
Finally, in the third step (Test$_{DG4}$) pulsating inflow conditions 
like those of case DG4 are imposed. Also, for this case, the emissivity 
field and synthetic PVDs are calculated. 
The test case results are reported in Fig.~\ref{Test_rhosmallbox},
Fig.~\ref{Test_emiss} and Fig.~\ref{Test_PVD}. \\
Fig.~\ref{Test_rhosmallbox} shows density and temperature fields
(top and bottom, respectively) for the test case (right coloumn) {\it vs.} 
the DG4 case (left coloumn). The patterns are basically the same. 
Actually, filaments of recirculating matter in the cocoon region, 
which are still visible in the DG4 case, have disappeared in Test$_{DG3}$,
because of the "older" age of the jet.
The panel of Fig.~\ref{Test_emiss} shows the emissivity patterns in 
the r-z plane for the DG4 case (on the left), and for the test case
(on the right), for both the stationary (top) and pulsated
(bottom) phase. The already mentioned filaments in the  
cocoon region do not provide relevant contribution to the jet 
emissivity, and play no role in PVDs formation, as shown by comparing 
Fig.~\ref{Test_PVD} with Fig.~\ref{pvd_05_lin} and Fig.~\ref{s2_non-steady}.\\
As a conclusion, the test case confirms that the adopted domain size 
and cooling model do not affect the bow-shock feedback in the emitting region
and the reliabilty of the results.

\end{document}